\documentclass{elegantpaper}
\usepackage[utf8]{inputenc}
\usepackage{authblk}
\usepackage{hyperref}
\usepackage{lettrine}

\makeatletter
\newcommand{\printfnsymbol}[1]{%
  \textsuperscript{\@fnsymbol{#1}}%
}
\makeatother

\title{Algorithmic Amplification of Politics on Twitter}

\author[a,b,c]{Ferenc Husz\'{a}r\thanks{equal contribution}\textsuperscript{,}\thanks{corresponding authors, email: fhuszar@twitter.com, lbelli@twitter.com}\textsuperscript{,}}
\author[a]{Sofia Ira Ktena\printfnsymbol{1}\textsuperscript{,}\thanks{now at DeepMind}\textsuperscript{,}} 
\author[a]{Conor O'Brien\printfnsymbol{1}\textsuperscript{,}}
\author[a]{Luca Belli\printfnsymbol{2}\textsuperscript{,}}
\author[a]{Andrew Schlaikjer}
\author[d]{Moritz Hardt\thanks{MH was a paid consultant at Twitter. Work performed while consulting for Twitter.}\textsuperscript{,}}

\affil[a]{Twitter, 1355 Market St San Francisco, CA 94103, USA.}
\affil[b]{Computer Laboratory, University of Cambridge, Cambridge, UK.}
\affil[c]{Gatsby Computational Neuroscience Unit, University College London, London, UK.}
\affil[d]{Department of Electrical Engineering and Computer Sciences, University of California, Berkeley, CA, USA.}

\date{}

\begin{document}

\maketitle

\begin{abstract}
Content on Twitter's home timeline is selected and ordered by personalization algorithms. By consistently ranking certain content higher, these algorithms may amplify some messages while reducing the visibility of others. There's been intense public and scholarly debate about the possibility that some political groups benefit more from algorithmic amplification than others. We provide quantitative evidence from a long-running, massive-scale randomized experiment on the Twitter platform that committed a randomized control group including nearly 2M daily active accounts to a reverse-chronological content feed free of algorithmic personalization. We present two sets of findings. First, we studied Tweets by elected legislators from major political parties in 7 countries. Our results reveal a remarkably consistent trend: In 6 out of 7 countries studied, the mainstream political right enjoys higher algorithmic amplification than the mainstream political left. Consistent with this overall trend, our second set of findings studying the U.S. media landscape revealed that algorithmic amplification favours right-leaning news sources. We further looked at whether algorithms amplify far-left and far-right political groups more than moderate ones: contrary to prevailing public belief, we did not find evidence to support this hypothesis. We hope our findings will contribute to an evidence-based debate on the role personalization algorithms play in shaping political content consumption.
\end{abstract}

Political content is a major part of the public conversation on Twitter. Politicians, political organizations, and news outlets engage large audiences on Twitter. At the same time, Twitter employs algorithms that learn from data to sort content on the platform. This interplay of algorithmic content curation and political discourse has been the subject of intense scholarly debate and public scrutiny~\cite{hong2016political, economist, nytimes, conover2011political, bail2018exposure, himelboim2013birds, hasson2020manipulators, barbera2015tweeting, o2015down,gillespie2010politics, bozdag2015breaking,ribeiro2020auditing, shirky2011political, parmelee2011politics, freelon2020false}.
When first established as a service, Twitter used to present individuals with content from accounts they followed, arranged in a reverse chronological feed. In 2016, Twitter introduced machine learning algorithms to render Tweets on this feed called Home timeline based on a personalized relevance model~\cite{rankingblogpost}. Individuals would now see older Tweets deemed relevant to them, as well as some Tweets from accounts they did not directly follow.

Personalized ranking prioritizes some Tweets over others on the basis of content features, social connectivity, and user activity. There is evidence that different political groups use Twitter differently to achieve political goals~\cite{badawy2018analyzing, conover2012partisan, vergeer2013online, graham2013between}. What has remained a matter of debate, however, is whether or not any ranking advantage falls along established political contours, such as, the left or right~\cite{economist, hasson2020manipulators}, the center or the extremes~\cite{hong2016political, nytimes}, specific parties~\cite{economist, hasson2020manipulators}, or news sources of a certain political inclination~\cite{bakshy2015exposure}. In this work, we provide the first systematic quantitative insights into this question based on a massive scale randomized experiment on the Twitter platform.

\section*{Experimental Setup}
Below we outline this experimental setup and its inherent limitations. We then introduce a measure of algorithmic amplification in order to quantify the degree to which different political groups benefit from algorithmic personalization.

When Twitter introduced machine learning to personalize the Home timeline in 2016, it excluded a randomly chosen control group of 1\% of all global Twitter users from the new personalized Home timeline. Individuals in this control group have never experienced personalized ranked timelines. Instead their Home timeline continues to display Tweets and Retweets from accounts they follow in reverse-chronological order. The treatment group corresponds to a sample of 4\% of all other accounts who experience the personalized Home timeline. However, even individuals in the treatment group do have the option to opt out of personalization (SI Section A).

The experimental setup has some inherent limitations. A first limitation stems from interaction effects between individuals in the analysis~\cite{aronow2017estimating}. In social networks, the control group can never be isolated from indirect effects of personalization as individuals in the control group encounter content shared by users in the treatment group. Therefore, although a randomized controlled experiment, our experiment does not satisfy the well-known Stable Unit Treatment Value Assumption (SUTVA) from causal inference~\cite{cox1958planning}. As a consequence, it cannot provide unbiased estimates of causal quantities of interest, such as the average treatment effect (ATE). In this study, we chose to not employ intricate causal inference machinery that is often used to approximate causal quantities~\cite{eckles2016design}, as these would not guarantee unbiased estimates in the complex setting of Twitter’s home timeline algorithm. Building an elaborate causal diagram of this complex system is well beyond the scope of our observational study. Instead, we present findings based on simple comparison of measurements between the treatment and control groups. Intuitively, we expect peer effects to decrease observable differences between the control and treatment groups, thus, our reported statistics likely underestimate the true causal effects of personalization.

A second limitation pertains to the fact that differences between treatment and control group were previously used by Twitter to improve the personalized ranking experience. The treatment, i.e., the ranking experience, has therefore not remained the same over time. Moreover, the changes to the treatment depend on the experiment itself.

\section*{Measuring Amplification}
We define the reach of a set \textit{T} of Tweets in a set \textit{U} of Twitter users as the total number of users from \textit{U} who encountered a Tweet from the set \textit{T}\,\footnote{A Tweet is counted as "encountered" by user A when 50\% of the UI element containing the Tweet is continuously visible on the user’s device for 500ms. See SI Materials and Methods for details.}. Think of \textit{T}, for example, as Tweets from a group of politicians in Germany and the audience \textit{U} as all German Twitter users in the control group. We always consider reach within a specific time window, e.g., a day.

\subsection*{Measuring Algorithmic Amplification}
We define the amplification ratio of set \textit{T} of Tweets in an audience \textit{U} as the ratio of the reach of \textit{T} in \textit{U} intersected with the treatment group and the reach of \textit{T} in \textit{U} intersected with the control group. We normalize the ratio in such a way that amplification ratio 0\% corresponds to equal proportional reach in treatment and control. In other words, a random user from \textit{U} in the treatment group is just as likely to see a Tweet in \textit{T} as a random user from \textit{U} in the control group. An amplification ratio of 50\% means that the treatment group is 50\% more likely to encounter one of the Tweets. Large amplification ratios indicate that the ranking model assigns higher relevance scores to the set of Tweets, which therefore appear more often than they would in a reverse-chronological ordering.

We often study the amplification ratio in cases where \textit{T} is a set corresponding to Tweets from a single Twitter account (individual amplification). When considering how groups of accounts are amplified we have the choice between reporting distribution of amplification ratios of the individual accounts in the group, or to consider a single aggregate amplification ratio (group amplification), where \textit{T} contains all Tweets authored by any member of the group. We generally report both statistics. More detail on how we calculate amplification and a discussion of the difference between individual and group amplification is found in Section D of SI.

\section*{Results}
We divide our findings into two parts. First, we study Tweets by elected politicians from major political parties in seven countries which were highly represented on the platform. In the second analysis, which is specific to the United States, we study whether algorithmic amplification of content from major media outlets is associated with political leaning.

We first report how personalization algorithms amplify content from elected officials from various political parties and parliamentary groups. We identified Twitter account details and party affiliation for currently serving legislators in 7 countries from public data~\cite{vrandevcic2014wikidata,mpsandlords, ballotpedia, canadaparl} (SI Section B). The countries in our analysis were chosen on the basis of data availability: these countries have a large enough active Twitter user base for our analysis, and it was possible to obtain details of legislators from high-quality public sources. In cases where a legislator has multiple accounts---for example, an official and a personal account---we included all of them in the analysis. In total we identified 3,634 accounts belonging to legislators across the 7 countries (the combined size of legislatures is 3,724 representatives). We then selected original Tweets authored by the legislators, including any replies and quote Tweets (where they retweet a Tweet while also adding original commentary). We excluded retweets without comment, as attribution is ambiguous when multiple legislators retweet the same content. When calculating amplification relating to legislators, we considered their reach only within their respective country during the time period between 1 April 2020 and 15 August 2020.

\begin{figure*}[t!]
\begin{align*}
\includegraphics[width=\textwidth,clip,trim={1.3cm 0.9cm 1.3cm 0.8cm}]{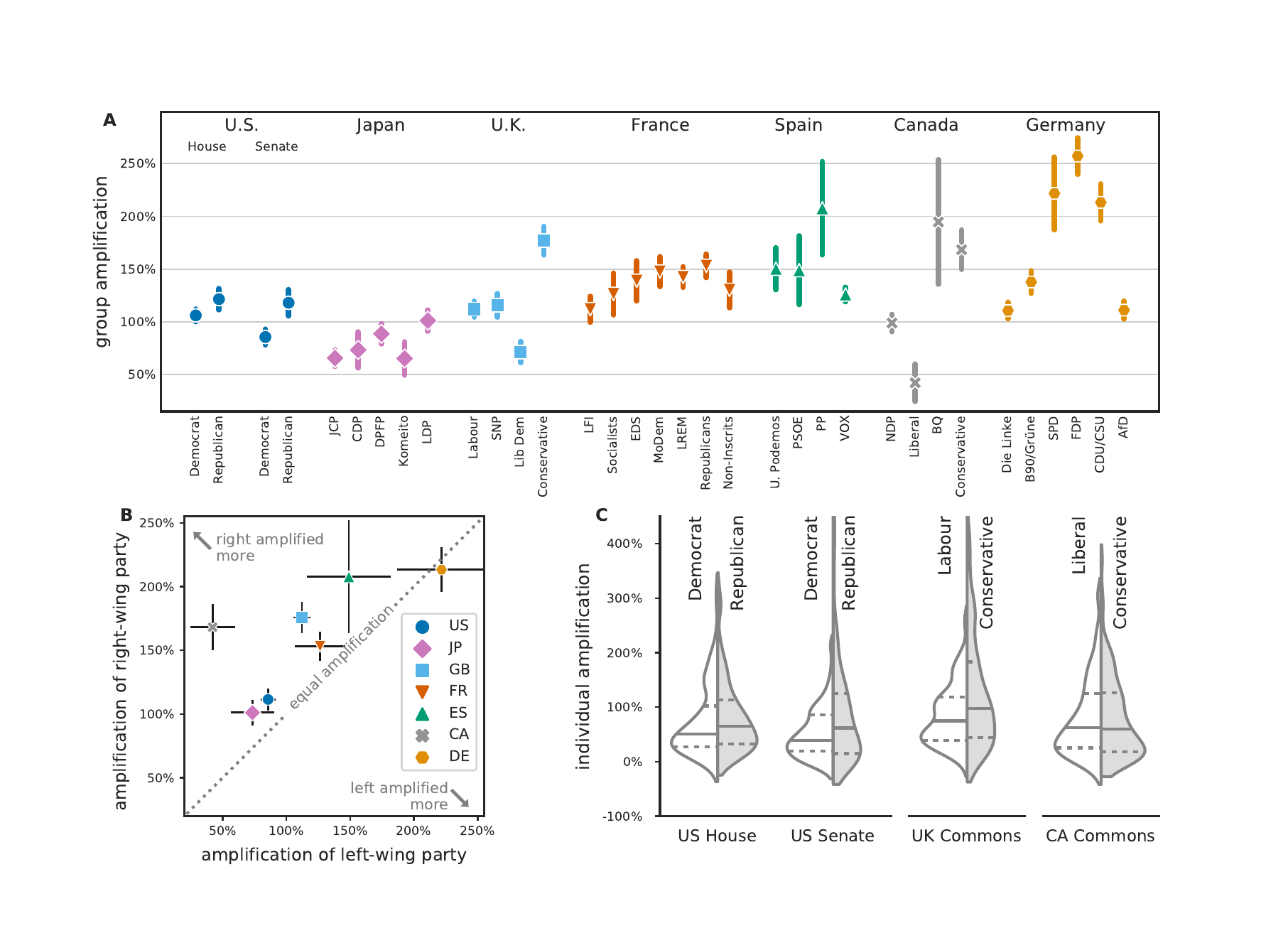}
\end{align*}
\caption{Amplification of Tweets from major political groups and politicians in 7 countries with an active Twitter user base. \textbf{A:} Group amplification of each political party or group. Within each country, parties are ordered from left to right according to their ideological position based on the 2019 Chapel Hill Expert Survey \cite{ches2019}. A value of $0\%$ indicates that Tweets by the group reach the same number of users on ranked timelines as they do on chronological timelines. A value of $100\%$ means double the reach. Error bars show standard error estimated from bootstrap. Bootstrap resampling was performed over daily intervals as well as membership of each political group. \textbf{B:} Pairwise comparison between the largest mainstream left and right-wing parties in each country: Democrats vs Republicans in the U.S., CDP vs LDP in Japan, Labour vs Conservatives in the U.K., Socialists vs Republicans in France, PSOE vs Popular in Spain, Liberals vs Conservatives in Canada and SPD vs CDU/CSU in Germany. In 6 out of 7 countries, these comparisons yield a statistically significant difference with right being amplified more, after adjusting for multiple comparisons. In Germany, the difference is not statistically significant. \textbf{C:} Amplification of Tweets by individual left- and right-wing politicians in the U.S., U.K. and Canada. Violin plots illustrate the distribution of amplification values within each party, solid and dashed lines within show the median, 25th and 75th percentiles, respectively. There is substantial variation of individual amplification within political parties. However, there is no statistically significant dependence between an individual’s amplification and their party affiliation in either of the four comparisons.}
\label{fig:electedofficials_combined}
\end{figure*}

To compare the amplification of political groups, we can either calculate the amplification of all Tweets from the group (group amplification, Fig.~\ref{fig:electedofficials_combined}A-B, or calculate amplification of each individual in the group separately (individual amplification, Fig.~\ref{fig:electedofficials_combined}C). The latter yields a distribution of individual amplification values for each group, thus revealing individual differences of amplifying effects within a group.

Fig. \ref{fig:electedofficials_combined}A compares the group amplification of major political parties in the countries we studied. Values over 0\% indicate that all parties enjoy an amplification effect by algorithmic personalization, in some cases exceeding 200\%, indicating that the party’s Tweets are exposed to an audience three times the size of the audience they reach on chronological timelines. To test the hypothesis that left-wing or right-wing politicians are amplified differently, we identified the largest mainstream left or centre-left and mainstream right or centre-right party in each legislature, and present pairwise comparisons between these in Fig.~\ref{fig:electedofficials_combined}B. With the exception of Germany, we find a statistically significant difference favoring the political right wing. This effect is strongest in Canada (Liberals 43\% vs Conservatives 167\%) and the United Kingdom (Labour 112\% vs Conservatives 176\%). In both countries the Prime Ministers and members of the Government are also Members of the Parliament and are thus included in our analysis. We therefore recomputed the amplification statistics after excluding top government officials. Our findings, shown in SI Fig.\ S2., remained qualitatively similar.

When studying amplification at the level of individual politicians (Fig.~\ref{fig:electedofficials_combined}C), we find that amplification varies substantially within each political party: while Tweets from some individual politicians are amplified up to 400\%, for others amplification is below 0\%, meaning they reach fewer users on ranked timelines than they do on chronological ones. We repeated the comparison between major left-wing and right-wing parties, comparing the distribution of individual amplification values between parties. When studied at the individual level, a permutation test detected no statistically significant association between an individual’s party affiliation and their amplification.

We see that comparing political parties on the basis of aggregate amplification of the entire party (Fig.\ 1A-B) or on the basis of individual amplification of their members (Fig.\ 1C) leads to seemingly different conclusions: while individual amplification is not associated with party membership, the aggregate group amplification may be different for each party. These findings are not contradictory, considering that different politicians may reach overlapping audiences. Even if the amplification of individual politicians is uncorrelated with their political affiliation, when we consider increases to their combined reach, group-level correlations might emerge. For a more detailed discussion please refer to SI Section 1.E.3.

Our fine-grained data also allows us to evaluate whether recommender systems amplify extreme ideologies, far-left or far-right politicians, over more moderate ones~\cite{nytimes}. We found that in countries where far-left or far-right parties have substantial representation among elected officials (e.g. VOX in Spain, Die Linke and AfD in Germany, LFI and RN in France) the amplification of these parties is generally lower than that of moderate/centrist parties in the same country (see Fig. S1). Finally, we considered whether personalization consistently amplifies messages from governing coalition or the opposition, and found no consistent pattern across countries. For example, in the United Kingdom amplification favors the governing Conservatives, while in Canada the opposition Conservative Party of Canada is more highly amplified.

\begin{figure*}[bt!]
\begin{align*}
\includegraphics[width=16.5cm,clip,trim={0cm 0.9cm 1.6cm 0cm}]{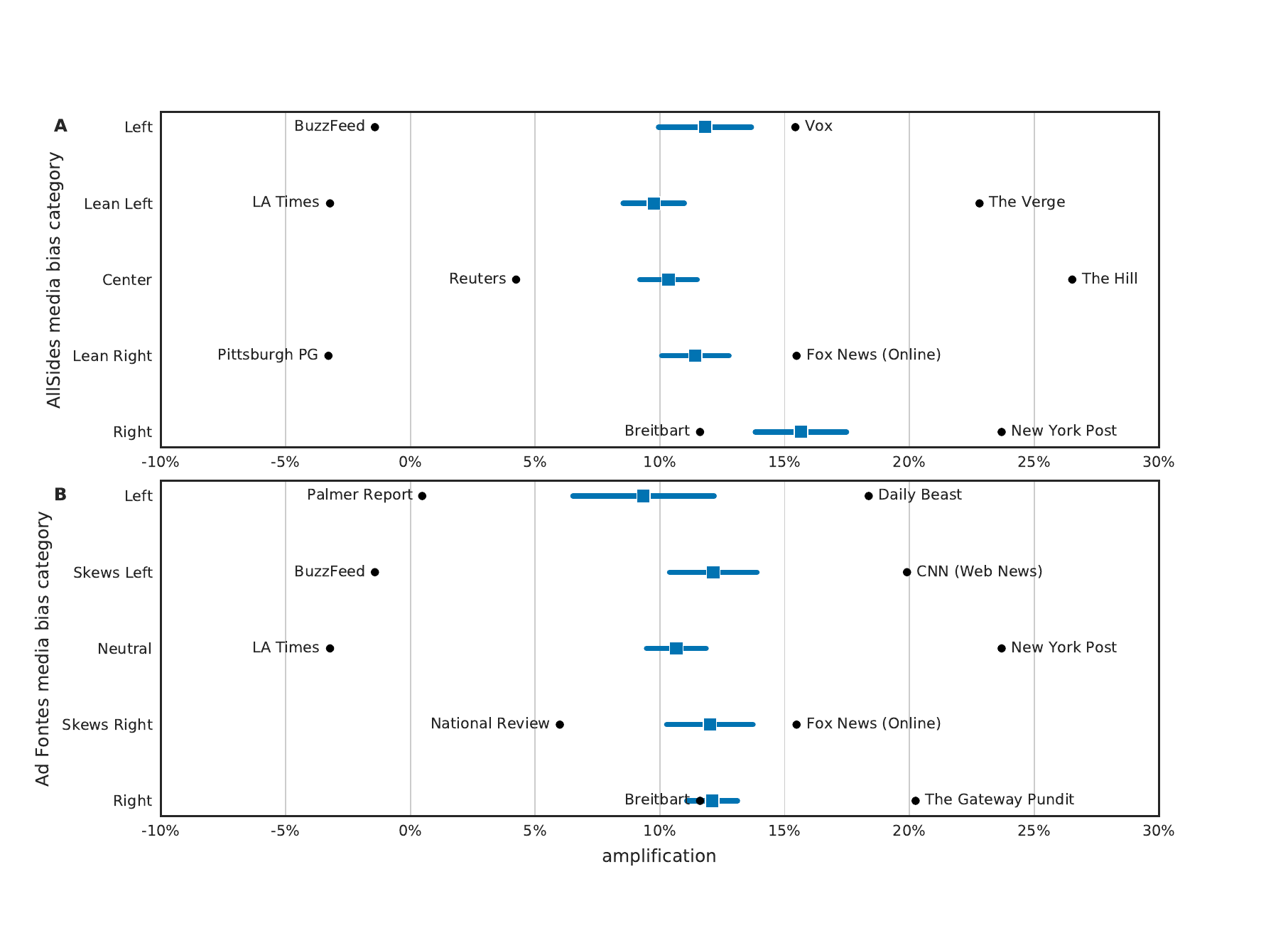}
\end{align*}
\caption{Amplification of news articles by Twitter’s personalization algorithms broken down by AllSides (\textbf{A}) and Ad Fontes (\textbf{B}) media bias ratings of their source. Blue squares denote the mean estimate of group amplification for each group of content, error bars show the standard deviation of the bootstrap estimate. Individual black circles show the amplification for the most significant positive and negative outliers within each group. For example, content from AllSides 'Left' media bias category is amplified $12\%$ by algorithms. The most significant negative outlier in this group is BuzzFeed, with an amplification of $-2\%$ compared to the chronological baseline. By contrast, Vox is amplified $16\%$. Negative and positive outliers are selected by a leave-one-out procedure detailed in SI Section 1.E.4.}
\label{fig:allsides_adfontes_results}
\end{figure*}

Tweets from legislators cover just a small portion of political content on the platform. To better understand the effects of personalization on political discourse, we extend our analysis to a broader domain of news content~\cite{messing2014selective,flaxman2016filter}. Specifically, we extend our analysis to media outlets with a significant audience in the U.S.~\cite{american2014personal}. While the political affiliation of a legislator is publicly verifiable, there is no single agreed-upon classification of the political orientation of media outlets.

To reduce subjectivity in our classification of political content, we leverage two independently curated media bias rating datasets from AllSides~\cite{allsides} and Ad Fontes Media\footnote{\href{https://www.adfontesmedia.com/interactive-media-bias-chart/}{Ad Fontes Media Bias Chart 5.0}}~\cite{adfontes}, and present results for both. Both datasets assign labels to media sources based on their perceived position on the U.S. media bias landscape. The labels describe the overall media bias of a news source on a 5-point scale ranging from partisan Left through Center/Neutral to partisan Right. We then identified Tweets containing links to articles from these news sources shared by anyone between 1 April 2020 and 15 August 2020. We excluded Tweets pointing to non-political content such as recipes or sports. Wherever possible, we separated editorial content from general news coverage as in some cases these had different bias ratings (SI Section 1.C). The resulting dataset contains AllSides annotations for 100,575,284 unique Tweets pointing to 6,258,032 articles and Ad Fontes annotations for 88,818,544 unique Tweets pointing to 5,100,381 articles.

We then grouped Tweets by media bias annotation of their source and calculated the aggregate amplification of each bias category (Fig.~\ref{fig:allsides_adfontes_results}) based on impressions over the time period between 15 April and 15 August 2020. When using AllSides bias ratings (Fig.~\ref{fig:allsides_adfontes_results}A), two general trends emerge: The personalization algorithms amplify more partisan sources compared to ones rated as Center. Secondly, the partisan Right is amplified marginally more compared to the partisan Left. The results based on Ad Fontes bias ratings (Fig.~\ref{fig:allsides_adfontes_results}B) differ in some key ways. Most notable is the relatively low, 10.5\%, amplification of the partisan Left compared to other categories. Among the remaining categories, the differences are not substantial, although the Neutral category is amplified significantly less than other categories.

Leave-one-out analysis of each media bias category (described in detail in SI Section 1.E.4) allows us to identify the most significant outliers in each category, also shown in Fig.~\ref{fig:allsides_adfontes_results}. This analysis identified BuzzFeed News, LA Times and Breitbart (based on both AllSides and Ad Fontes ratings) as negative outliers in their respective categories, meaning the amplification of their content was less than the aggregate amplification of the bias category they belong to. Meanwhile, Fox News and New York Post were identified as positive outliers. These outliers also illustrate that, just as we saw in the case of legislators, there is significant variation among news outlets in each bias category.

The fact that our findings differ depending on the media bias dataset used underlines the critical reliance of this type of analysis on political labels. We do not endorse either AllSides or Ad Fontes as objectively better ratings, and leave it to the reader to interpret the findings according to their own assessment. To aid this interpretation, we looked at how AllSides and Ad Fontes ratings differ, where both ratings are available. We found that while the two rating schemes largely agree on rating the political right, they differ most in their assessment of publications on the political left, with a tendency for Ad Fontes to rate publications as being more neutral compared to their corresponding AllSides rating. Details are shown in Supplementary Figs.\ S3, S4 and Table S1.

\section*{Discussion}
We presented a comprehensive audit of algorithmic amplification of political content by the recommender system in Twitter’s home timeline. Across seven countries we studied, we found that mainstream right-wing parties benefit at least as much, and often substantially more, from algorithmic personalization as their left-wing counterparts. In agreement with this, we found that content from U.S. media outlets with a strong right-leaning bias are amplified marginally more than content from left-leaning sources. However, when making comparisons based on the amplification of individual politician’s accounts, rather than parties in aggregate, we found no association between amplification and party membership.

Our analysis of far-left and far-right parties in various countries does not support the hypothesis that algorithmic personalization amplifies extreme ideologies more than mainstream political voices. However, some findings point at the possibility that strong partisan bias in news reporting is associated with higher amplification. We note that strong partisan bias here means a consistent tendency to report news in a way favouring one party or another, and does not imply the promotion of extreme political ideology.

Recent arguments that different political parties pursue different strategies on Twitter~\cite{parmelee2011politics, freelon2020false} may provide an explanation as to why these disparities exist. However, understanding the precise causal mechanism that drives amplification invites further study that we hope our work initiates.

Although it is the first systematic and large-scale study contrasting ranked timelines with chronological ones on Twitter, our work fits into a broader context of research on the effects of content personalization on political content~\cite{economist, nytimes,o2015down, bakshy2015exposure} and polarization~\cite{perra2019modelling, pariser2011filter, del2016echo,boxell2017greater}. There are several avenues for future work. Apart from the Home timeline, Twitter users are exposed to several other forms of algorithmic content curation on the platform that merit study through similar experiments. Political amplification is only one concern with online recommendations. A similar methodology may provide insights into domains such as misinformation~\cite{del2016spreading, grinberg2019fake}, manipulation~\cite{woolley2016automating,aral2019protecting}, hate speech and abusive content.

\bibliography{holdback}

\begin{thebibliography}{}

\bibitem[{AllSides}, 2020]{allsides}
{AllSides} (2020).
\newblock {How AllSides Rates Media Bias}.
\newblock
  [\href{https://www.allsides.com/media-bias/media-bias-rating-methods}{Online};
  retrieved August-2020].

\bibitem[Aral and Eckles, 2019]{aral2019protecting}
Aral, S. and Eckles, D. (2019).
\newblock Protecting elections from social media manipulation.
\newblock {\em Science}, 365(6456):858--861.

\bibitem[Aronow et~al., 2017]{aronow2017estimating}
Aronow, P.~M., Samii, C., et~al. (2017).
\newblock Estimating average causal effects under general interference, with
  application to a social network experiment.
\newblock {\em The Annals of Applied Statistics}, 11(4):1912--1947.

\bibitem[Badawy et~al., 2018]{badawy2018analyzing}
Badawy, A., Ferrara, E., and Lerman, K. (2018).
\newblock Analyzing the digital traces of political manipulation: The 2016
  russian interference twitter campaign.
\newblock In {\em ACM Int. Conf. on ASONAM}, pages 258--265. IEEE.

\bibitem[Bail et~al., 2018]{bail2018exposure}
Bail, C.~A., Argyle, L.~P., Brown, T.~W., Bumpus, J.~P., Chen, H., Hunzaker,
  M.~F., Lee, J., Mann, M., Merhout, F., and Volfovsky, A. (2018).
\newblock Exposure to opposing views on social media can increase political
  polarization.
\newblock {\em Proceedings of the National Academy of Sciences},
  115(37):9216--9221.

\bibitem[Bakker et~al., 2020]{ches2019}
Bakker, R., Hooghe, L., Jolly, S., Marks, G., Polk, J., Ronvy, J., Steenbergen,
  M., and Vachudova, M.~A. (2020).
\newblock 2019 {Chapel Hill Expert Survey}, version 2019.1., available on
  \href{http://chesdata.eu}{chesdata.eu}.

\bibitem[Bakshy et~al., 2015]{bakshy2015exposure}
Bakshy, E., Messing, S., and Adamic, L.~A. (2015).
\newblock Exposure to ideologically diverse news and opinion on facebook.
\newblock {\em Science}, 348(6239):1130--1132.

\bibitem[Ballotpedia, 2020]{ballotpedia}
Ballotpedia (2020).
\newblock {List of Current Members of US Congress.}
\newblock
  [\href{https://ballotpedia.org/List_of_current_members_of_the_U.S._Congress}{Online};
  accessed June-2020].

\bibitem[Barber{\'a} et~al., 2015]{barbera2015tweeting}
Barber{\'a}, P., Jost, J.~T., Nagler, J., Tucker, J.~A., and Bonneau, R.
  (2015).
\newblock Tweeting from left to right: Is online political communication more
  than an echo chamber?
\newblock {\em Psychological science}, 26(10):1531--1542.

\bibitem[Boxell et~al., 2017]{boxell2017greater}
Boxell, L., Gentzkow, M., and Shapiro, J.~M. (2017).
\newblock Greater internet use is not associated with faster growth in
  political polarization among us demographic groups.
\newblock {\em Proceedings of the National Academy of Sciences},
  114(40):10612--10617.

\bibitem[Bozdag and van~den Hoven, 2015]{bozdag2015breaking}
Bozdag, E. and van~den Hoven, J. (2015).
\newblock Breaking the filter bubble: democracy and design.
\newblock {\em Ethics and Information Technology}, 17(4):249--265.

\bibitem[Conover et~al., 2012]{conover2012partisan}
Conover, M.~D., Gon{\c{c}}alves, B., Flammini, A., and Menczer, F. (2012).
\newblock Partisan asymmetries in online political activity.
\newblock {\em EPJ Data Science}, 1(1):6.

\bibitem[Conover et~al., 2011]{conover2011political}
Conover, M.~D., Ratkiewicz, J., Francisco, M.~R., Gon{\c{c}}alves, B., Menczer,
  F., and Flammini, A. (2011).
\newblock Political polarization on twitter.
\newblock {\em ICWSM}, 133(26):89--96.

\bibitem[Cox, 1958]{cox1958planning}
Cox, D.~R. (1958).
\newblock {\em Planning of experiments.}
\newblock Wiley.

\bibitem[Del~Vicario et~al., 2016a]{del2016spreading}
Del~Vicario, M., Bessi, A., Zollo, F., Petroni, F., Scala, A., Caldarelli, G.,
  Stanley, H.~E., and Quattrociocchi, W. (2016a).
\newblock The spreading of misinformation online.
\newblock {\em Proceedings of the National Academy of Sciences},
  113(3):554--559.

\bibitem[Del~Vicario et~al., 2016b]{del2016echo}
Del~Vicario, M., Vivaldo, G., Bessi, A., Zollo, F., Scala, A., Caldarelli, G.,
  and Quattrociocchi, W. (2016b).
\newblock Echo chambers: Emotional contagion and group polarization on
  facebook.
\newblock {\em Scientific reports}, 6:37825.

\bibitem[Eckles et~al., 2016]{eckles2016design}
Eckles, D., Karrer, B., and Ugander, J. (2016).
\newblock Design and analysis of experiments in networks: Reducing bias from
  interference.
\newblock {\em Journal of Causal Inference}, 5(1).

\bibitem[Flaxman et~al., 2016]{flaxman2016filter}
Flaxman, S., Goel, S., and Rao, J.~M. (2016).
\newblock Filter bubbles, echo chambers, and online news consumption.
\newblock {\em Public opinion quarterly}, 80(S1):298--320.

\bibitem[Freelon et~al., 2020]{freelon2020false}
Freelon, D., Marwick, A., and Kreiss, D. (2020).
\newblock False equivalencies: Online activism from left to right.
\newblock {\em Science}, 369(6508):1197--1201.

\bibitem[Gillespie, 2010]{gillespie2010politics}
Gillespie, T. (2010).
\newblock The politics of ‘platforms’.
\newblock {\em New media \& society}, 12(3):347--364.

\bibitem[Graham et~al., 2013]{graham2013between}
Graham, T., Broersma, M., Hazelhoff, K., and Van'T~Haar, G. (2013).
\newblock Between broadcasting political messages and interacting with voters:
  The use of twitter during the 2010 uk general election campaign.
\newblock {\em Information, communication \& society}, 16(5):692--716.

\bibitem[Grinberg et~al., 2019]{grinberg2019fake}
Grinberg, N., Joseph, K., Friedland, L., Swire-Thompson, B., and Lazer, D.
  (2019).
\newblock Fake news on twitter during the 2016 us presidential election.
\newblock {\em Science}, 363(6425):374--378.

\bibitem[Hasson, 2020]{hasson2020manipulators}
Hasson, P.~J. (2020).
\newblock {\em The Manipulators: Facebook, Google, Twitter, and Big Tech's War
  on Conservatives}.
\newblock Regnery Publishing.

\bibitem[Himelboim et~al., 2013]{himelboim2013birds}
Himelboim, I., McCreery, S., and Smith, M. (2013).
\newblock Birds of a feather tweet together: Integrating network and content
  analyses to examine cross-ideology exposure on twitter.
\newblock {\em Journal of computer-mediated communication}, 18(2):154--174.

\bibitem[Hong and Kim, 2016]{hong2016political}
Hong, S. and Kim, S.~H. (2016).
\newblock Political polarization on twitter: Implications for the use of social
  media in digital governments.
\newblock {\em Government Information Quarterly}, 33(4):777--782.

\bibitem[{House of Commons Canada}, 2020]{canadaparl}
{House of Commons Canada} (2020).
\newblock {Current Members of Parliament.}
\newblock [\href{https://www.ourcommons.ca/members/en/search}{Online}; accessed
  June-2020].

\bibitem[Institute, 2014]{american2014personal}
Institute, A.~P. (2014).
\newblock The personal news cycle: How americans choose to get their news.
\newblock
  [\href{https://www.americanpressinstitute.org/publications/reports/survey-research/personal-news-cycle/}{Online};
  retrieved October-2020].

\bibitem[Messing and Westwood, 2014]{messing2014selective}
Messing, S. and Westwood, S.~J. (2014).
\newblock Selective exposure in the age of social media: Endorsements trump
  partisan source affiliation when selecting news online.
\newblock {\em Communication research}, 41(8):1042--1063.

\bibitem[Otoro, 2020]{adfontes}
Otoro, V. (2020).
\newblock {How Ad Fontes Ranks News Sources - Methodology Summary.}
\newblock
  [\href{https://www.adfontesmedia.com/how-ad-fontes-ranks-news-sources/}{Online};
  retrieved August-2020].

\bibitem[O’Callaghan et~al., 2015]{o2015down}
O’Callaghan, D., Greene, D., Conway, M., Carthy, J., and Cunningham, P.
  (2015).
\newblock Down the (white) rabbit hole: The extreme right and online
  recommender systems.
\newblock {\em Social Science Computer Review}, 33(4):459--478.

\bibitem[Pariser, 2011]{pariser2011filter}
Pariser, E. (2011).
\newblock {\em The filter bubble: What the Internet is hiding from you}.
\newblock Penguin UK.

\bibitem[Parmelee and Bichard, 2011]{parmelee2011politics}
Parmelee, J.~H. and Bichard, S.~L. (2011).
\newblock {\em Politics and the Twitter revolution: How tweets influence the
  relationship between political leaders and the public}.
\newblock Lexington Books.

\bibitem[Perra and Rocha, 2019]{perra2019modelling}
Perra, N. and Rocha, L.~E. (2019).
\newblock Modelling opinion dynamics in the age of algorithmic personalisation.
\newblock {\em Scientific reports}, 9(1):1--11.

\bibitem[Ribeiro et~al., 2020]{ribeiro2020auditing}
Ribeiro, M.~H., Ottoni, R., West, R., Almeida, V.~A., and Meira~Jr, W. (2020).
\newblock Auditing radicalization pathways on {YouTube}.
\newblock In {\em Proceedings of FAT*}, pages 131--141.

\bibitem[Shirky, 2011]{shirky2011political}
Shirky, C. (2011).
\newblock The political power of social media: Technology, the public sphere,
  and political change.
\newblock {\em Foreign Affairs}, 90(1):28--41.

\bibitem[{The Economist}, 2020]{economist}
{The Economist} (2020).
\newblock Twitter’s algorithm does not seem to silence conservatives.
\newblock
  [\href{https://www.economist.com/graphic-detail/2020/08/01/twitters-algorithm-does-not-seem-to-silence-conservatives}{Online};
  posted 01-August-2020].

\bibitem[Tufekci, 2018]{nytimes}
Tufekci, Z. (2018).
\newblock Youtube, the great radicalizer.
\newblock
  [\href{https://www.nytimes.com/2018/03/10/opinion/sunday/youtube-politics-radical.html}{Online};
  posted 10-March-2018].

\bibitem[Twitter, 2016]{rankingblogpost}
Twitter (2016).
\newblock Never miss important tweets from people you follow.
\newblock
  [\href{https://www.economist.com/graphic-detail/2020/08/01/twitters-algorithm-does-not-seem-to-silence-conservatives}{Online};
  accessed 01-August-2020].

\bibitem[{UK Parliament}, 2020]{mpsandlords}
{UK Parliament} (2020).
\newblock {MPs and Lords.}
\newblock [\href{https://members.parliament.uk/members/Commons}{Online};
  accessed June-2020].

\bibitem[Vergeer et~al., 2013]{vergeer2013online}
Vergeer, M., Hermans, L., and Sams, S. (2013).
\newblock Online social networks and micro-blogging in political campaigning:
  The exploration of a new campaign tool and a new campaign style.
\newblock {\em Party politics}, 19(3):477--501.

\bibitem[Vrande{\v{c}}i{\'c} and Kr{\"o}tzsch, 2014]{vrandevcic2014wikidata}
Vrande{\v{c}}i{\'c}, D. and Kr{\"o}tzsch, M. (2014).
\newblock Wikidata: a free collaborative knowledgebase.
\newblock {\em Communications of the ACM}, 57(10):78--85.

\bibitem[Woolley, 2016]{woolley2016automating}
Woolley, S.~C. (2016).
\newblock Automating power: Social bot interference in global politics.
\newblock {\em First Monday}, 21(4).

\end{thebibliography}


\begin{thebibliography}{}

\bibitem[{AllSides}, 2020]{allsides}
{AllSides} (2020).
\newblock {How AllSides Rates Media Bias}.
\newblock
  [\href{https://www.allsides.com/media-bias/media-bias-rating-methods}{Online};
  retrieved August-2020].

\bibitem[Bakker et~al., 2020]{ches2019}
Bakker, R., Hooghe, L., Jolly, S., Marks, G., Polk, J., Ronvy, J., Steenbergen,
  M., and Vachudova, M.~A. (2020).
\newblock 2019 {Chapel Hill Expert Survey}, version 2019.1., available on
  \href{http://chesdata.eu}{chesdata.eu}.

\bibitem[Equy, 2020]{lremeds}
Equy, L. (2020).
\newblock Ressac de nœuds pour lrem.
\newblock {\em Liberation}.

\bibitem[Otoro, 2020]{adfontes}
Otoro, V. (2020).
\newblock {How Ad Fontes Ranks News Sources - Methodology Summary.}
\newblock
  [\href{https://www.adfontesmedia.com/how-ad-fontes-ranks-news-sources/}{Online};
  retrieved August-2020].

\bibitem[{Twitter}, 2020]{earningsq22020}
{Twitter} (2020).
\newblock Q2 2020 letter to shareholders.
\newblock published by Twitter Investor Relations,
  [\href{https://s22.q4cdn.com/826641620/files/doc_financials/2020/q2/Q2-2020-Shareholder-Letter.pdf}{Online};
  retrieved August-2020].

\bibitem[Vrande{\v{c}}i{\'c} and Kr{\"o}tzsch, 2014]{vrandevcic2014Wikidata}
Vrande{\v{c}}i{\'c}, D. and Kr{\"o}tzsch, M. (2014).
\newblock Wikidata: a free collaborative knowledgebase.
\newblock {\em Communications of the ACM}, 57(10):78--85.

\end{thebibliography}

\end{document}

\maketitle

\section{Materials and Methods}

\subsection{The Timelines Quality Holdback Experiment}

Twitter has maintained the randomized experiment described in this paper, known as the timelines quality holdback experiment, since June 2016. The experiment allocates accounts to either a control (1\% of users) or treatment (4\%) group completely at random.

\subsubsection{Assignment to Treatment or Control}
Accounts were randomly assigned to treatment or control either at the experiment’s onset (if the account existed at the time), or at the time the account was created (if the account was created since the experiment started). In both cases inclusion in the experiment and assignment to treatment or control were determined completely randomly. The assignment is maintained over the lifespan of the account, although users in the treatment group have the liberty to temporarily switch off algorithmic recommendations (as explained below). For the purposes of this study, individuals who voluntarily turn off algorithmic recommendations continue to be identified as being in the treatment group.

\subsubsection{Number of users in the Experiment}
The experiment included 5\% of all accounts globally, of which 20\% are assigned to control, and 80\% are assigned to the treatment group. This amounts to tens of millions of unique user ids including those of dormant accounts and bots, only a fraction of which correspond to active users who used the product during the time period we studied. In the second quarter of 2020, Twitter reported 186 million monetizable daily active users (mDAU) \cite{earningsq22020} of which about 5\% (9.3 million) were included in our study.

\subsubsection{Tweet Selection and Presentation in Control Group}
Users assigned to the control group experience the Twitter home timeline much the same way it worked before the introduction of algorithmic ranking in 2016~\footnote{\url{https://help.twitter.com/en/using-twitter/twitter-timeline}}. The timeline displays Tweets authored by accounts the individual follows (referred to as in-network Tweets) or Tweets reTweeted by accounts the individual follows (ReTweets). Tweets are ordered in reverse chronological order, showing the latest Tweet first, taking into account the timestamp of either Tweet creation (for original Tweets) or reTweet time (for ReTweets). A heuristic de-duplication logic is applied to Tweets which are reTweeted by multiple accounts the individuals follow. Due to the simplicity of Tweet selection and ranking logic, this version of the timeline is often referred to as the non-algorithmic timeline.

\subsubsection{Tweet Selection and Ranking in Treatment Group}
Users in the treatment group for this experiment experience the Twitter Home Timeline the same way that the majority of Twitter users - not included in the experiment - do. Being included in the treatment bucket of this experiment has an influence on whether the same user can be selected as part of the treatment group in other randomized experiments in which related features are tested.

Treatment users who use the Twitter app on iOS or Android devices can choose between viewing the “top Tweets first” or the “latest Tweets first” in their Home timeline\footnote{\url{https://blog.twitter.com/official/en_us/a/2016/never-miss-important-Tweets-from-people-you-follow.html}}. By default, the “top Tweets first” option is enabled and the setting reverts back to this default after a period of inactivity. On other platforms, users do not have an option to view latest Tweets first. When the Home timeline is set to display ‘latest Tweets first’, it works similarly to the traditional reverse chronological timeline which users in the control group experience.
When ‘top Tweets first’ mode is selected, the Tweets are algorithmically selected, filtered and ranked. The selection and ranking of Tweets is influenced, in part, by the output of machine learning models which are trained to predict whether the user is likely to engage with the Tweet in various ways (like, reTweet, reply, etc). The machine learning methods and ranking algorithms make these predictions based on a combination of content signals such as the inferred topic of the Tweet as well as behavioural signals such as past engagement history between the user and the author of the Tweet.

In addition to Tweets and ReTweets from accounts the individual follows, the personalized timeline may also contain ‘injected’ content from outside the user’s immediate network. A common type of such injected content are Tweets liked by someone the individual follows. Such Tweets would not appear on chronological timelines, only if someone the user follows ReTweets (rather than just likes) the Tweet in question.

\subsubsection{Services influencing Content in both Control and Treatment}
Several services and products might influence content displayed on both Control and Treatment Timelines.

\textit{Promoted Tweets}: The timeline might contain Promoted Tweets, a form of advertisement, which are marked by a label “Promoted by [advertiser account]” making them clearly distinguishable from organic content. These Tweets are not usually from accounts the individual follows. These promoted Tweets are selected by algorithms which rely on machine learning algorithms and a generalized second price auction mechanism. Although the ad models might behave differently for ranking and control users,  the algorithms selecting content have no explicit knowledge of whether a user is in either group.

\textit{Restricted or Removed content}: content deemed to violate Twitter’s Terms and Conditions, pornographic content, content displaying gore and violence, spam and fraudulent content, Tweets containing misleading or false information might be completely hidden, displayed with a warning label, or remain hidden until the user dismisses a warning message. These limitations tend to apply to individual Tweets or all Tweets from certain Twitter accounts, and apply the same way every time the Tweet.

\textit{Blocking and Muting}: Users have the ability to Mute or Block one another. Individuals who Mute another account will no longer see Tweets from the muted account, even when it is reTweeted or replied to by someone they follow. Blocking works the other way around, if an individual is blocked by an account, they can no longer see or interact with content from that account.

\subsubsection{Machine learning models used for ranking}
The ranking content on the Home timeline is influenced by the output of deep learning models, trained to predict various types of engagements with Tweets (likes, reTweets, replies, etc)~\footnote{\href{https://blog.twitter.com/engineering/en_us/topics/insights/2017/using-deep-learning-at-scale-in-twitters-timelines.html}{Using Deep Learning at Scale in Twitter’s Timelines.}}. The models receive as input various signals which broadly fall into the following categories: content features (such as the inferred topic area of the Tweet, whether it contains an image), about the engaging user (such as their engagement history or users they follow), about the Tweet author (such as engagements with their past Tweets), and about the relationship between the user and the Tweet’s author (such as frequency of past engagements between them). In addition to the predictions made by these machine learning models, the final ranking of Tweets is influenced by heuristics and the output of other machine learning algorithms.

The models are trained on data from users in this experiment’s treatment group as well as most users outside of this experiment (the remaining 95\% of accounts). Data from individuals in the control group are not part of the training dataset of these models. However, the predictions the algorithms make are routinely evaluated on data from control users, and this is used to inform decisions about the deployment of certain features.

\subsubsection{Other forms of personalization users are exposed to}
Being assigned to the control or treatment group of this experiment only affects algorithmic personalization of the Home timeline. Users receive content recommendations or are served personalized results in other product surface areas including the “Explore” tab, Search, Trends, Events, push notifications and Tweet or conversation detail pages. In addition to this, users might receive personalized content recommendations via email or push notifications, as well as ‘Who to follow’ account recommendations. Therefore, it is incorrect to say that the control group received no algorithmic recommendations. Assignment to control only restricts the use of personalization in the Home timeline, which is the component users spend most active minutes in.

\subsection{Obtaining Legislators’ Twitter details \label{sec:legislators}}

\subsubsection{Selection criteria for countries \label{sec:countries_selection}}
When studying amplification of content from national legislators, our aim was to include data relating to as many legislatures globally as possible. We identified countries to include in our analysis based on the availability of data, in particular based on the following criteria:

\textit{Availability of data on politicians’ Twitter accounts}: It was possible for us to identify Twitter usernames and associated party affiliations for a large number of current legislators using either official government sources, Wikidata~\cite{vrandevcic2014Wikidata}, or publicly available Twitter lists curated by the corresponding political parties or other official accounts. We used the following \href{https://query.wikidata.org/#%23Find%20parliaments%20with%20the%20largest%20number%20of%20Twitter%20handles%20associated%20with%20MPs.%0ASELECT%0A%20%20%20%20%28%3FposHeldLabel%20AS%20%3Fparliament%29%0A%20%20%20%20%28COUNT%28DISTINCT%20LCASE%28%3FtwitterScreenname%29%29%20AS%20%3FtwitterUsers%29%0AWHERE%20%7B%0A%20%20%3Fperson%20wdt%3AP39%20%3FposHeld%20.%0A%20%20%3Fperson%20wdt%3AP2002%20%3FtwitterScreenname%20.%0A%23%20%20%3FposHeldStatement%20ps%3AP39%20%3FposHeld%20.%0A%20%20%20%20%20%20%20%20%20%20%20%20%20%20%23%20%20%20%20%20%20pq%3AP2937%20wd%3AQ18479094%20.%0A%20%20%3FposHeld%20wdt%3AP279%2B%20wd%3AQ486839%20.%0A%20%20SERVICE%20wikibase%3Alabel%20%7B%20bd%3AserviceParam%20wikibase%3Alanguage%20%22en%22.%20%0A%20%20%20%20%20%20%20%20%20%20%20%20%20%20%20%20%20%20%20%20%20%20%20%20%20%20%20%3FposHeld%20rdfs%3Alabel%20%3FposHeldLabel%0A%20%20%20%20%20%20%20%20%20%20%20%20%20%20%20%20%20%20%20%20%20%20%20%20%20%20%7D%0A%7D%0AGROUP%20BY%20%3FposHeldLabel%0AORDER%20BY%20DESC%28%3FtwitterUsers%29}{Wikidata query} to list the number of Twitter accounts associated with each legislature.

\textit{Sufficient Twitter user base in the country}: The number of unique Twitter accounts in the control group which had the relevant country code associated with their account at the time of entering the experiment was at least 100,000 supporting robust statistical analysis.

Screening for the above two criteria we identified the following list of countries (in decreasing order by number of unique users in our experiment) United States, Japan, United Kingdom, France, Spain, Canada, Germany, and Turkey. Further analysis of Wikidata entries for legislators in Turkey showed that while there was almost complete data on legislators serving in the previous, 26th term, for the current, 27th term, we were only able to identify 110 usernames for a total of 600 members. Thus we did not include Turkey in our analysis.
The following countries met the first, but not the second selection criterion, and were thus not included: Sweden, Denmark and the Netherlands. The following countries met the second, but not the first requirement and were thus not included: India, Brazil, Mexico, Saudi Arabia, Indonesia.

We excluded federal or international bodies such as the European Union Parliament from our analysis and focussed on national legislatures only. Members of the EU legislature Tweet in different languages and address audiences in different countries whose activity and usage of Twitter may not be homogeneous enough, complicating the analysis and interpretation of findings.

\subsubsection{Collecting data from Wikidata}
To identify members of the current legislative term in each country we followed a variation of the following process:
We identified the appropriate subclass of the ‘legislator’ Wikidata entity (Q4175034) which describes membership in each legislature we studied. For example, the Wikidata entity Q3044918 denotes the position “member of the French National Assembly”. These entities linked to the Wikidata entity of each person who have held the title via the Wikidata property P39. “position held”. This allows us to identify Wikidata entities of all current and past members of the legislature.

In countries where such Wikidata entities are available, we identified the entity describing the current term of legislature. These entities are linked to the title held via the Wikidata qualifier P2937. For example the Wikidata entity describing the 15th, current, term of the French legislature is Q24939798. These entities can then be used to identify current members of the legislature.

We additionally discarded individuals whose membership of the corresponding legislature had an “end date” (P582) qualifier, as this indicates that the person no longer holds the position. In some countries, like the United States, this is the only way to find current members of the Senate as there are no Wikidata entities for legislative terms.
Where available, we can retrieve legislators’ Twitter screen names which are linked to the individual’s Wikidata entity via the P2002 Wikidata property. In some cases, it was also possible to retrieve numerical Twitter identifiers. Where available, numerical IDs are more reliable as users can change their screen names. We found that legislators change their twitter screennames quite often, for example, to include “mp” or “MdB” signifying their membership of Parliament or the German Bundestag as they are elected.

We also retrieved individuals’ gender or sex, which is linked to a person’s entity via the Wikidata property P21. We only used gender information to ensure that members of all genders are fairly represented in our sample of Twitter accounts.

Examples of SQPARQL queries we used in each country can be accessed here: \href{https://query.wikidata.org/#%23%23%23%23%23%0A%23Members%20of%20Bundestag%20with%20Twitter%20accounts%0A%23%23%23%23%23%0ASELECT%0A%20%20%20%20%28%22DE%22%20AS%20%3FcountryCode%29%0A%20%20%20%20%28%3FpersonLabel%20AS%20%3Fname%29%0A%20%20%20%20%28%3Fperson%20AS%20%3FwikidataID%29%0A%20%20%20%20%28%3FgenderLabel%20AS%20%3Fgender%29%0A%20%20%20%20%28GROUP_CONCAT%28DISTINCT%20LCASE%28%3FtwitterScreenname%29%3B%20SEPARATOR%3D%22%7C%7C%22%29%20AS%20%3FtwitterScreenname%29%0A%20%20%20%20%28GROUP_CONCAT%28DISTINCT%20%3FposHeldLabel%3B%20SEPARATOR%3D%22%7C%7C%22%29%20AS%20%3FpositionHeld%29%0A%20%20%20%20%28GROUP_CONCAT%28DISTINCT%20%3FgroupLabel%3B%20SEPARATOR%3D%22%7C%7C%22%29%20AS%20%3Fgroup%29%0AWHERE%20%7B%0A%20%20%23%20retrieving%20positions%20held%2C%20and%20statement%20properties%0A%20%20%3Fperson%20p%3AP39%20%3FposHeldStatement.%0A%20%20%3FposHeldStatement%20ps%3AP39%20%3FposHeld%20.%0A%0A%20%20%23%20%27member%20of%20the%20German%20Bundestag%27%0A%20%20%7B%3FposHeld%20wdt%3AP279%2a%20wd%3AQ1939555%20.%7D%0A%20%20%0A%20%20%23current%20member%2C%2019th%20Bundsestag%0A%20%20%3FposHeldStatement%20pq%3AP2937%20wd%3AQ30579723%20.%0A%20%20%0A%20%20%23retrieving%20gender%20if%20specified%0A%20%20OPTIONAL%20%7B%3Fperson%20wdt%3AP21%20%3Fgender%20.%7D%0A%20%20%0A%20%20%23retrieving%20Twitter%20username%20if%20specified%0A%20%20OPTIONAL%20%7B%3Fperson%20wdt%3AP2002%20%3FtwitterScreenname%20.%7D%0A%20%20%0A%20%20%23retrieving%20group%20if%20specified%0A%20%20OPTIONAL%20%7B%3FposHeldStatement%20pq%3AP4100%20%3Fgroup%20.%7D%0A%0A%20%20%23adding%20human%20readable%20labels%0A%20%20SERVICE%20wikibase%3Alabel%20%7B%20bd%3AserviceParam%20wikibase%3Alanguage%20%22en%22.%20%0A%20%20%20%20%20%20%20%20%20%20%20%20%20%20%20%20%20%20%20%20%20%20%20%20%20%20%3FposHeld%20rdfs%3Alabel%20%3FposHeldLabel.%0A%20%20%20%20%20%20%20%20%20%20%20%20%20%20%20%20%20%20%20%20%20%20%20%20%20%20%3Fgroup%20rdfs%3Alabel%20%3FgroupLabel.%0A%20%20%20%20%20%20%20%20%20%20%20%20%20%20%20%20%20%20%20%20%20%20%20%20%20%20%3Fperson%20rdfs%3Alabel%20%3FpersonLabel.%0A%20%20%20%20%20%20%20%20%20%20%20%20%20%20%20%20%20%20%20%20%20%20%20%20%20%20%3Fgender%20rdfs%3Alabel%20%3FgenderLabel.%0A%20%20%20%20%20%20%20%20%20%20%20%20%20%20%20%20%20%20%20%20%20%20%20%20%20%7D%0A%7D%0AGROUP%20BY%20%3Fperson%20%3FpersonLabel%20%3FgenderLabel%0AORDER%20BY%20%3FpositionHeld}{Germany}, \href{https://query.wikidata.org/#%23%23%23%23%23%0A%23Catalan%20representatives%20with%20Twitter%20accounts%0A%23%23%23%23%23%0ASELECT%0A%20%20%20%20%28%22ES%22%20AS%20%3FcountryCode%29%0A%20%20%20%20%28%3FpersonLabel%20AS%20%3Fname%29%0A%20%20%20%20%28%3Fperson%20AS%20%3FwikidataID%29%0A%20%20%20%20%28%3FgenderLabel%20AS%20%3Fgender%29%0A%20%20%20%20%28GROUP_CONCAT%28DISTINCT%20LCASE%28%3FtwitterScreenname%29%3B%20SEPARATOR%3D%22%7C%7C%22%29%20AS%20%3FtwitterScreenname%29%0A%20%20%20%20%28GROUP_CONCAT%28DISTINCT%20%3FposHeldLabel%3B%20SEPARATOR%3D%22%7C%7C%22%29%20AS%20%3FpositionHeld%29%0A%20%20%20%20%28GROUP_CONCAT%28DISTINCT%20%3FgroupLabel%3B%20SEPARATOR%3D%22%7C%7C%22%29%20AS%20%3Fgroup%29%0AWHERE%20%7B%0A%20%20%23%20retrieving%20positions%20held%2C%20and%20statement%20properties%0A%20%20%3Fperson%20p%3AP39%20%3FposHeldStatement.%0A%20%20%3FposHeldStatement%20ps%3AP39%20%3FposHeld%20.%0A%0A%20%20%0A%20%20%23%20selecting%20legistrators%20in%20Spain%0A%20%20%23%20%27member%20of%20Parliament%20of%20Catalonia%27%0A%20%20%7B%3FposHeld%20wdt%3AP279%2a%20wd%3AQ18714088%20.%7D%0A%20%20UNION%20%23or%0A%20%20%23%20%27member%20of%20Congress%20of%20Deputies%20of%20Spain%27%0A%20%20%7B%3FposHeld%20wdt%3AP279%2a%20wd%3AQ18171345%20.%7D%0A%20%20%0A%20%20%23selecting%20current%20legislators%0A%20%20%20%23current%20member%2C%2012th%20Legislature%20in%20Catalonia%0A%20%20%7B%3FposHeldStatement%20pq%3AP2937%20wd%3AQ47034616%20.%7D%0A%20%20UNION%20%23or%0A%20%20%23current%20member%2C%2014th%20Legislature%20in%20Spain%0A%20%20%7B%3FposHeldStatement%20pq%3AP2937%20wd%3AQ77871368%20.%7D%20%0A%20%20%0A%20%20%23retrieving%20gender%20if%20specified%0A%20%20OPTIONAL%20%7B%3Fperson%20wdt%3AP21%20%3Fgender%20.%7D%0A%20%20%0A%20%20%23retrieving%20Twitter%20username%20if%20specified%0A%20%20OPTIONAL%20%7B%3Fperson%20wdt%3AP2002%20%3FtwitterScreenname%20.%7D%0A%20%20%0A%20%20%23retrieving%20group%20if%20specified%0A%20%20OPTIONAL%20%7B%3FposHeldStatement%20pq%3AP4100%20%3Fgroup%20.%7D%0A%0A%20%20%23adding%20human%20readable%20labels%0A%20%20SERVICE%20wikibase%3Alabel%20%7B%20bd%3AserviceParam%20wikibase%3Alanguage%20%22en%22.%20%0A%20%20%20%20%20%20%20%20%20%20%20%20%20%20%20%20%20%20%20%20%20%20%20%20%20%20%3FposHeld%20rdfs%3Alabel%20%3FposHeldLabel.%0A%20%20%20%20%20%20%20%20%20%20%20%20%20%20%20%20%20%20%20%20%20%20%20%20%20%20%3Fgroup%20rdfs%3Alabel%20%3FgroupLabel.%0A%20%20%20%20%20%20%20%20%20%20%20%20%20%20%20%20%20%20%20%20%20%20%20%20%20%20%3Fperson%20rdfs%3Alabel%20%3FpersonLabel.%0A%20%20%20%20%20%20%20%20%20%20%20%20%20%20%20%20%20%20%20%20%20%20%20%20%20%20%3Fgender%20rdfs%3Alabel%20%3FgenderLabel.%0A%20%20%20%20%20%20%20%20%20%20%20%20%20%20%20%20%20%20%20%20%20%20%20%20%20%7D%0A%7D%0AGROUP%20BY%20%3Fperson%20%3FpersonLabel%20%3FgenderLabel%0AORDER%20BY%20%3FpositionHeld}{Spain}, \href{https://query.wikidata.org/#%23French%20deputies%20in%20the%20National%20Assembly%20with%20Twitter%20account%20and%20gender%0ASELECT%0A%20%20%20%20%28%22FR%22%20AS%20%3FcounryCode%29%0A%20%20%20%20%28%3FpersonLabel%20AS%20%3Fname%29%0A%20%20%20%20%28%3Fperson%20AS%20%3FwikidataID%29%0A%20%20%20%20%28%3FgenderLabel%20AS%20%3Fgender%29%0A%20%20%20%20%28GROUP_CONCAT%28DISTINCT%20LCASE%28%3FtwitterScreenname%29%3B%20SEPARATOR%3D%22%7C%7C%22%29%20AS%20%3FtwitterScreenname%29%0A%20%20%20%20%28GROUP_CONCAT%28DISTINCT%20%3FposHeldLabel%3B%20SEPARATOR%3D%22%7C%7C%22%29%20AS%20%3FpositionHeld%29%0A%20%20%20%20%28GROUP_CONCAT%28DISTINCT%20%3FgroupLabel%3B%20SEPARATOR%3D%22%7C%7C%22%29%20AS%20%3Fgroup%29%0AWHERE%20%7B%0A%20%20%23%20getting%20positions%20held%2C%20and%20the%20statement%20properties%0A%20%20%3Fperson%20p%3AP39%20%3FposHeldStatement.%0A%20%20%3FposHeldStatement%20ps%3AP39%20%3FposHeld%20%3B%0A%20%20%20%20%20%20%20%20%20%20%20%20%20%20%20%20%20%20%20%20pq%3AP2937%20wd%3AQ24939798%20.%20%20%20%20%20%20%20%20%20%20%20%20%20%20%23current%20member%2C%20i.e.%20position%20annotated%20by%20current%20legislative%20term%0A%20%20OPTIONAL%20%7B%3FposHeldStatement%20pq%3AP4100%20%3Fgroup%20.%7D%20%20%20%20%20%20%20%20%23retrieving%20coalition%20if%20specified%0A%20%20%0A%20%20%23restricting%20to%20Japanese%20House%20of%20Representatives%0A%20%20%3FposHeld%20wdt%3AP279%2a%20wd%3AQ3044918%20.%20%20%20%20%20%20%20%20%20%20%20%20%20%20%20%20%20%20%20%20%20%20%23position%20is%20a%20subclass%20of%20%27Member%20of%20French%20National%20Assembly%27%0A%20%20%0A%20%20%23retrieving%20gender%20if%20available%0A%20%20OPTIONAL%20%7B%3Fperson%20wdt%3AP21%20%3Fgender%7D%20.%0A%20%20%23retrieving%20Twitter%20username%20if%20available%0A%20%20OPTIONAL%20%7B%3Fperson%20wdt%3AP2002%20%3FtwitterScreenname%7D%20.%0A%0A%20%20SERVICE%20wikibase%3Alabel%20%7B%20bd%3AserviceParam%20wikibase%3Alanguage%20%22en%22.%20%0A%20%20%20%20%20%20%20%20%20%20%20%20%20%20%20%20%20%20%20%20%20%20%20%20%20%20%3FposHeld%20rdfs%3Alabel%20%3FposHeldLabel.%0A%20%20%20%20%20%20%20%20%20%20%20%20%20%20%20%20%20%20%20%20%20%20%20%20%20%20%3Fgroup%20rdfs%3Alabel%20%3FgroupLabel.%0A%20%20%20%20%20%20%20%20%20%20%20%20%20%20%20%20%20%20%20%20%20%20%20%20%20%20%3Fperson%20rdfs%3Alabel%20%3FpersonLabel.%0A%20%20%20%20%20%20%20%20%20%20%20%20%20%20%20%20%20%20%20%20%20%20%20%20%20%20%3Fgender%20rdfs%3Alabel%20%3FgenderLabel.%0A%20%20%20%20%20%20%20%20%20%20%20%20%20%20%20%20%20%20%20%20%20%20%20%20%20%7D%0A%7D%0AGROUP%20BY%20%3Fperson%20%3FpersonLabel%20%3FgenderLabel}{France} and \href{https://query.wikidata.org/#%23%23%23%23%23%0A%23Members%20of%20Swedish%20Riksdag%20with%20Twitter%20accounts%0A%23%23%23%23%23%0ASELECT%0A%20%20%20%20%28%22SW%22%20AS%20%3FcountryCode%29%0A%20%20%20%20%28%3FpersonLabel%20AS%20%3Fname%29%0A%20%20%20%20%28%3Fperson%20AS%20%3FwikidataID%29%0A%20%20%20%20%28%3FgenderLabel%20AS%20%3Fgender%29%0A%20%20%20%20%28GROUP_CONCAT%28DISTINCT%20LCASE%28%3FtwitterScreenname%29%3B%20SEPARATOR%3D%22%7C%7C%22%29%20AS%20%3FtwitterScreenname%29%0A%20%20%20%20%28GROUP_CONCAT%28DISTINCT%20%3FtwitterIdentifier%3B%20SEPARATOR%3D%22%7C%7C%22%29%20AS%20%3FtwitterID%29%0A%20%20%20%20%28GROUP_CONCAT%28DISTINCT%20%3FposHeldLabel%3B%20SEPARATOR%3D%22%7C%7C%22%29%20AS%20%3FpositionHeld%29%0A%20%20%20%20%28GROUP_CONCAT%28DISTINCT%20%3FgroupLabel%3B%20SEPARATOR%3D%22%7C%7C%22%29%20AS%20%3Fgroup%29%0AWHERE%20%7B%0A%20%20%23%20retrieving%20positions%20held%2C%20and%20statement%20properties%0A%20%20%3Fperson%20p%3AP39%20%3FposHeldStatement.%0A%20%20%3FposHeldStatement%20ps%3AP39%20%3FposHeld%20.%0A%0A%20%20%23%20%27member%20of%20the%20Swedish%20Rikstag%27%0A%20%20%7B%3FposHeld%20wdt%3AP279%2a%20wd%3AQ10655178%20.%7D%0A%20%20%0A%20%20%23current%20member%2C%20i.e.%20no%20end%20date%20specified%0A%20%20MINUS%20%7B%20%3FposHeldStatement%20pq%3AP582%20%3FendTime.%20%7D%0A%20%20%3FposHeldStatement%20pq%3AP2937%20wd%3AQ76961916%20.%0A%0A%20%20%23retrieving%20gender%20if%20specified%0A%20%20OPTIONAL%20%7B%3Fperson%20wdt%3AP21%20%3Fgender%20.%7D%0A%20%20%0A%20%20%23retrieving%20Twitter%20screenname%20if%20specified%0A%20%20OPTIONAL%20%7B%3Fperson%20wdt%3AP2002%20%3FtwitterScreenname%20.%7D%0A%20%20%0A%20%20%23retrieving%20Twitter%20userid%20if%20specified%0A%20%20OPTIONAL%20%7B%20%3Fperson%20p%3AP2002%2Fpq%3AP6552%20%3FtwitterIdentifier%20.%20%7D%0A%20%20%0A%20%20%23retrieving%20group%20if%20specified%0A%20%20OPTIONAL%20%7B%3FposHeldStatement%20pq%3AP4100%20%3Fgroup%20.%7D%0A%0A%20%20%23adding%20human%20readable%20labels%0A%20%20SERVICE%20wikibase%3Alabel%20%7B%20bd%3AserviceParam%20wikibase%3Alanguage%20%22en%22.%20%0A%20%20%20%20%20%20%20%20%20%20%20%20%20%20%20%20%20%20%20%20%20%20%20%20%20%20%3FposHeld%20rdfs%3Alabel%20%3FposHeldLabel.%0A%20%20%20%20%20%20%20%20%20%20%20%20%20%20%20%20%20%20%20%20%20%20%20%20%20%20%3Fgroup%20rdfs%3Alabel%20%3FgroupLabel.%0A%20%20%20%20%20%20%20%20%20%20%20%20%20%20%20%20%20%20%20%20%20%20%20%20%20%20%3Fperson%20rdfs%3Alabel%20%3FpersonLabel.%0A%20%20%20%20%20%20%20%20%20%20%20%20%20%20%20%20%20%20%20%20%20%20%20%20%20%20%3Fgender%20rdfs%3Alabel%20%3FgenderLabel.%0A%20%20%20%20%20%20%20%20%20%20%20%20%20%20%20%20%20%20%20%20%20%20%20%20%20%7D%0A%7D%0AGROUP%20BY%20%3Fperson%20%3FpersonLabel%20%3FgenderLabel%0AORDER%20BY%20%3FpositionHeld}{Sweden}.

\subsubsection{Collecting data from public Twitter lists}
In addition to Wikidata, we have obtained lists of Twitter accounts of legislators in the United States and Germany from publicly available Twitter lists curated by Twitter (@TwitterGov) or political parties and political organizations in Germany (\href{https://twitter.com/i/lists/63915247}{US House}, \href{https://twitter.com/i/lists/63915645}{US Senate}, \href{https://twitter.com/i/lists/917896954008539136}{Bundestagsabgeordnete}, \href{https://twitter.com/i/lists/101365010}{CSU MdBs}, \href{https://twitter.com/i/lists/12971627}{MdBs 19.WP}, \href{https://twitter.com/i/lists/912252646953758720}{MdB DIE LINKE 19. WP}). We manually verified that the lists we chose were of high quality, accurate and up-to-date. We removed accounts which belong to parties, local party organisations or campaign groups rather than individual legislators.

\subsubsection{Collecting data from official government websites}
To obtain a high-quality list of Twitter handles for Members of the United Kingdom Parliament we scraped the official Parliament website (\href{https://members.parliament.uk/members/Commons}{UK Parliament: MPs and Lords}). The contact page of each MP may contain their Twitter handle, as well as their party affiliation. For the United States we also accessed a list of representatives and senators from Ballotpedia (\href{https://ballotpedia.org/List_of_current_members_of_the_U.S._Congress. Accessed June 2020}{Ballotpedia “List of Current Members of US Congress”}). We used this list to associate the accounts found on the Twitter lists to their political parties. The Ballotpedia information was also used to identify a handful of accounts that were not already included in the curated lists. For Canada, a CSV of current MPs was downloaded from the official House of Commons website (\href{https://www.ourcommons.ca/members/en/search}{House of Commons Canada: Current Members of Parliament}). The accounts were then manually annotated resulting in a very high quality dataset in terms of coverage and accuracy.

\subsubsection{Manual data validation and annotation}
While in most countries we were able to identify Twitter details of over 70\% of all representatives following the automated methods described above, and in some countries this coverage was particularly high, we cannot be certain that we did not miss individuals. Our goal was to ensure that when we miss accounts, these do not result in poor representation of certain minority groups in our dataset. We have therefore focussed manual annotation effort on ensuring that accounts of legislators who belong to certain underrepresented groups, such as women, and people of colour, are included in our dataset. In most countries, we were able to retrieve gender labels from Wikidata to aid with his process.

\subsubsection{Groupings of parties}

To test various hypotheses about the types of political parties algorithms might amplify more, we make some direct comparisons between parties in each country. The first comparison we present in Fig.~\ref{fig:party_amplification}. A compares the mainstream political left with the mainstream political right. We used a number of heuristics to determine how to make this comparison in each country. In all countries except France, the parties being compared have the largest representation in their corresponding legislature. We rely on the 2019 Chapel Hill Expert Survey \cite{ches2019} and Wikidata annotations to determine the ideological position of each party. Typically, Wikipedia and other public sources describe one of these parties as left-wing or centre-left, and the other as right-wing or centre-right, making the comparison unambiguous. In France, the largest parliamentary group, the governing LREM is a centrist, big-tent parliamentary group, while the parties more traditionally considered as France’s left-wing and right-wing (Socialists and Republicans) are currently both in opposition. In Continental European countries, the left-wing parties compared (SPD, Parti Socialiste and PSOE) are all members of the Progressive Alliance of Socialists and Democrats in the European Parliament. Likewise, the right-wing parties (CDU/CSU, les R\'{e}publicains and Partido Popular) are part of the European People’s Party.

In Fig.~\ref{fig:party_amplification}B we compare far-left and far-right political parties with mainstream political parties from the same country. We selected political parties where Wikipedia entries mentioned an association with far-left or far-right ideologies, or where the 2019 Chapel Hill Expert Survey \cite{ches2019} indicated an extreme ideology (above 9 or below 2). These were the Japanese Communist Party (Japan far-left), La France insoumise (France far-left), Rassemblement National (France far-right, represented together with associates as Non-Inscrits in the French National Assembly), VOX (Spain, far-right), Die Linke (Germany far-left) and AfD (Germany far-right). We compared each of these parties or parliamentary groups to the largest mainstream right-wing or left-wing political party in their respective countries.

In Fig.~\ref{fig:party_amplification}C we compare governing vs opposition parties. In the United States, we consider Republicans to be the governing party and Democrats (and democratic-aligned independents like Bernie Sanders) as opposition. In Japan, the government is formed by LDP and Komeito, all other parties are considered opposition. In the United Kingdom we compare the Conservative Party against the Labour Party, according to their official designation as Her Majesty’s Government and Her Majesty’s Most Loyal Opposition, respectively. In France we consider LREM as well as their confidence-and-supply partners MoDem and Agir as the government, and all other representatives except EDS, which Wikipedia lists as neutral, as opposition. In Spain we consider PSOE, Unidas Podemos and their supporting  Basque Nationalist Party as governing, and Partido Popular, ERC and VOX as opposition. In Germany we consider CSU/CDU as governing and everyone else as opposition. In Canada we compare the Liberal Party against the Conservative Party according to their official designation as Her Majesty’s Government and Her Majesty’s Loyal Opposition, respectively.

\subsubsection{Political changes}
It is common for individual legislators’ party or parliamentary group affiliation changes during a legislative term. This can be due to individual resignations, party mergers or the formation of new parliamentary groups or parties. For example, the French Écologie Democratie Solidarité (EDS) parliamentary group was formed in May 2020 by members of the governing La République En Marche! (LREM)~\cite{lremeds}. In the Japanese House of Representatives, the Party of Hope (Kibō no Tō) merged with the Democratic Party to form a new party called DPFP in 2018, the majority of DPFP representatives then joined the Constitutional Democratic Party (CDP) in September 2020, while a smaller DPFP continues to exist. Where possible we validated and updated party membership data so as to best reflect major parties throughout the study period (1 April - 15 August 2020). In France, we considered EDS as separate from LREM. We repeated our analysis considering EDS representatives as members LREM, but findings do not change qualitatively. As can be seen in Fig.~\ref{fig:party_amplification}A, EDS and LREM are very similar in terms of group amplification.

\subsection{Media bias ratings  \label{sec:media_bias}}

\subsubsection{AllSides Media Bias Ratings}
To study exposure to politically biased media sources we obtained media bias ratings for news sources from AllSides~\cite{allsides}. While the AllSides dataset includes news sources with a global audience (such as the BBC, Guardian, Al Jazeera, etc) it focuses primarily on the U.S. media landscape, and the media bias ratings relate to how the media bias of these sources are perceived in the United States. This dataset assigns each publication or media source into one of ‘Left’, ‘Lean Left’, ‘Center’, ‘Mixed’, ‘Lean Right’ and ‘Right’. The data we had also contained crowdsourced judgments as well as confidence ratings which were ignored. We discarded the ‘Mixed’ category as it included aggregator websites and media bias rating platforms (like \url{AllSides.com} itself) rather than media sites creating original content. We have further excluded sites like Yahoo News, Google News, as well as podcasts, content studios and activist groups whose original content was not clearly identifiable or attributable. To qualify for our analysis, the content from the publication source had to be clearly identifiable on the basis of URLs shared by users on the platform, we discarded publications without a clearly identifiable URL structure.

\subsubsection{Ad Fontes Media Bias ratings}
We also obtained Media Bias ratings from Ad Fontes Media~\cite{adfontes}. This dataset contained a numerical media bias rating for each news source ranging between -38.5 (most extreme left bias) and 38.5 (most extreme right bias). In line with how the data is presented on the media bias chart~\footnote{\url{https://www.adfontesmedia.com/interactive-media-bias-chart/}} we discretized these numerical values into 5 intervals: Left ($x < -16.5$), Skews Left ($-16.5 < x < -5.5$), Neutral ($-5.5 < x < 5.5$), Skews Right ($5.5 < x < 16.5$) and Right ($16.5 < x$). We excluded news sources such as TV channels or programs whose content was not clearly identifiable as URLs shared on Twitter. We found that, with a few exceptions, the set of publications with an Ad Fontes rating was a subset of those with an AllSides rating.

\subsubsection{Mapping domain names and URLs}
The AllSides dataset contains websites (domain names) as well as details of Twitter accounts (usernames) associated with most publications they rated. One could therefore study exposure to Tweets from the official Twitter accounts (i.e. Tweets by @foxnews or @nytimes) or exposure to Tweets containing a link to content from each publication (i.e. Tweets containing a link to a Fox News of New York Times article). We chose to focus on URLs, because publications use their official Twitter accounts in different ways, and because Tweets from these official accounts are responsible for only a small fraction of all content impressions on Twitter. Thus, we chose to use domain names as the primary identifier of each publication. We manually added missing domain names and updated or corrected outdated entries to ensure that all major news sources had an up-to-date domain name.

\subsubsection{Regular expressions \label{sec:regex}}
To identify URLs that link to articles from each publication, we created regular expressions , which were matched against the text of the URL (not the text of the website). For the majority of publications these regular expressions were based on the domain name only, catching any link to any content on the corresponding website.
Several news sources publish large volumes of non-politicised content such as sports, recipes, games or weather forecasts. To filter these out and focus our analysis on news and politicised content only, we created custom regular expressions, which select only articles from certain sections of each publication only.

While the naming of sections and the URL template changed from publication to publication, we aimed at including sections which most likely contained coverage of local and global news, global health, COVID-19, politics, elections, climate, science and technology. Similarly, we excluded sports, wellness, food, or weather related sections.
Custom regular expressions also allowed us to distinguish between editorial and news articles from the same publisher. These often had different AllSides media bias ratings for the same newspaper group. For example, in the AllSides dataset, the online news from Fox News is rated ‘Lean Right’ while Fox News opinion pieces are rated ‘Right’. Similarly, the New York Times news section is rated ‘Lean Left’ while New York Times opinion section is rated ‘Left’. For publications where editorial content was rated differently, we tried to identify editorial content based on the URL string, using regular expressions. Unfortunately, this was not always possible. For example, all URLs for the Wall Street Journal were of the form `wsj.com/articles/[a-z0-9-]+`, making it impossible to identify opinion pieces based on the URL text only. In such cases, we did not distinguish between Editorial and News content, used all articles from the domain in our analysis, and assigned them to the AllSides rating of the non-editorial content (which tended to be less partisan).

We then identified Tweets with content from these publications by screening public Tweets created between 1 April and 15 August 2020, and matching any URLs these Tweets contained against the regular expressions we curated.  The resulting dataset contained AllSides annotations for 100,575,284 unique Tweets pointing to 6,258,032 different articles and Ad Fontes annotations for 88,818,544 unique Tweets pointing to 5,100,381 different articles. When calculating amplification on this data, we considered impressions within the time period 15 April -- 15 August 2020.

\subsubsection{Comparison of AllSides and Ad Fontes ratings}
To aid the interpretation of data, we have compared AllSides and Ad Fontes ratings for sources where we had both ratings available. Figure~\ref{fig:news_sources_count} shows the number of news sources for each pairing of Ad Fontes and AllSides rating. Figure~\ref{fig:news_sources_impressions} shows the breakdown in terms of number of Tweet impressions. Table~\ref{tab:ad_fontes_allsides_media} gives examples of specific news sources that have a certain combination of Ad Fontes and AllSides rating.

\subsection{Measuring amplification}

\subsubsection{Impression events}
Our measures of amplification are based on counting events called “linger impression”: these events are registered every time at least 50\% of the area of a Tweet is visible for at least 500 ms (including while scrolling). We note that these impression events are different from what is often called a ‘render impression’ in the context of online media, which is registered every time the Tweet is rendered in the user’s client, or every time a Tweet is fetched by the client on the user’s device. For example, when the user scrolls through a large volume of Tweets in rapid succession, without stopping to allow time to read or see any of the Tweets, several render impressions would be registered, while few or no linger impressions are logged. Linger impressions are the best proxy available to us to tell if a user has been exposed to the content of a Tweet. We note that this definition of a linger impression is not the authors' choice, and since it is hard-coded into the client software, it was not possible for us to consider alternative definitions of impression.

For the purposes of this paper we considered linger impressions registered on Android, iOS, desktop web clients. The timeline’s behaviour is implemented similarly in these platforms, and together they represent the overwhelming majority of user activity. We only considered linger impressions registered in the Home timeline component (the same users might encounter Tweets in other product areas such as Search, Trends, etc, but these impressions are not considered here). For each linger impression a country code is inferred from the user’s IP address at the time the event is registered. When analysing content from legislators, we further restrict impressions from the relevant country (for example when considering impressions of Tweets of French legislators, we only count impressions from France). When analysing content from publishers included in the AllSides media bias ratings, we restrict impression events from the United States.

\subsubsection{Defining amplification}

Let $T$ denote a set of Tweets. Let $U_{control}$ and $U_{treatment}$ denote the control and treatment group of users in the experiment, respectively. Note that, in our experiment, $|U_{treatment}|=4|U_{control}|$.
Let $U_{t,d}$ denote the set of users who registered a linger impression with Tweet $t$ on day $d$. For a set of Tweets $T$, we further define $U_{T,d} = \bigcup_{t \in T} U_{t,d}$, the set of users who encountered at least one Tweet from $T$ on day $d$. We define the amplification of the set of Tweets $T$ on day $d$ as:
$$a_d(T)= (\frac{|U_{T,d} \cap U_{treatment}| + 1}{4|U_{T,d} \cap U_{control}|+ 4} - 1) \cdot 100\%$$
Often, we consider amplification within a specific country. In this case we use $U_{t,d,c}$ in place of $U_{t,d}$, where $U_{t,d,c}$ denotes the set of all users who have registered an impression event involving Tweet $t$ while using Twitter from an IP address that we identified to be within the country $c$.

This gives rise to the following definition of amplification within country $c$ on day $d$:
$$a_{d,c}(T)= (\frac{|U_{T,d,c} \cap U_{treatment}| + 1}{4|U_{T,d,c} \cap U_{control}|+ 4} - 1) \cdot 100\%$$ 

For brevity, we will ignore the subscripts $c$ and $d$ in the discussions that follow, but note that amplification is always calculated on a daily basis, and within a specific country.
When we talk about the amplification of a group $G$ of individuals, such as members of a political party, we mean the amplification of the set of all Tweets created by this group $T_G$. The amplification for a group of users G is therefore:
$$a(G) = a(T_G)$$

Analogously, when referring to the amplification of an individual user $i$, we calculate this based on the set of Tweets, $T_i$, or equivalently, the group amplification for the singular set containing only $i$, that is $a(i) = a(\{i\})$.

When evaluating the amplification of news media, we consider the group of Tweets $T$ which contain a URL link that we identify to originate from a particular source, or a group of news sources, and calculate the amplification of this group of Tweets.

\subsection{Comparing amplification between groups of individuals}
Let's say we would like to compare two groups, $G_1$ and $G_2$, of users in terms of whether amplification favours group $G_1$ or $G_2$. Based on the amplification metrics we defined above we have two main ways of defining “equal” amplification.

\subsubsection{Equal group amplification} Since we have a way to measure the amplification of a group $G$ of individuals, expressed in terms of the amplification for the set of Tweets $T_G$, we can simply say the two groups are equally amplified if:
\begin{equation}
    \bar{a}(G_1) = \bar{a}(G2)
\end{equation}
where $\bar{a}$ is the average daily amplification. In our paper we calculate amplification on a daily basis over a 90 day period.
Note that this comparison is not robust to outliers within groups $G_1$ and $G_2$. If the groups contain very active individuals whose Tweets are seen more than others, their data would dominate the measure of amplification for the group. To increase the robustness to outliers we propose the following criterion:
Let $\tilde{G}_1$ and $\tilde{G}_2$ be bootstrap resampled variants of $G_1$ and $G_2$ respectively. That is, $\tilde{G}_1$ is a random subset of $G_1$ formed by sampling uniformly from $G_1$, $|G_1|$ times. We can say that group $G_1$ and $G_2$ are amplified equally if the following condition holds:
\begin{equation}
P[\bar{a}(\tilde{G}_1)>\bar{a}(\tilde{G}_2)] = P[\bar{a}(\tilde{G}_1)<\bar{a}(\tilde{G}_2)]\label{eqn:equal_group_amplification}
\end{equation}

It is possible to estimate $P[\bar{a}(\tilde{G}_1)>\bar{a}(\tilde{G}_2)]$ from data, and to determine whether it is significantly different from $P[\bar{a}(\tilde{G}_1)<\bar{a}(\tilde{G}_2)]$ by a permutation test.

\subsubsection{Individual amplification parity}
Another option is to calculate the amplification $\bar{a}_i$ for any individual $i$ in group $G_1$ and $G_2$ and compare the distribution of amplification values between the two groups. A strong criterion would require statistical independence between amplification and group membership, requiring the distribution of individual amplification values to completely agree between the groups. This is a very strong requirement and it is difficult to reliably establish independence when groups are small. Instead, we use a weaker notion of equivalence as follows. If $I_1$ is a random individual sampled uniformly from $G_1$ and $I_2$ is a random individual sampled uniformly from $G_2$ we say that the amplification of individuals in $G_1$ and $G_2$ is essentially equal if:
\begin{equation}
P[\bar{a}(I_1)>\bar{a}(I_2)] = P[\bar{a}(I_1)<\bar{a}(I_2)]
\label{eqn:individual_amplification_parity}
\end{equation}

Similarly to the bootstrap-based criterion for equal group amplification, we can estimate $P[\bar{a}(I_1)>\bar{a}(I_2)]$ from data, and determine whether it is significantly different from $P[\bar{a}(I_1)<\bar{a}(I_2)]$ using a permutation test.

\subsubsection{Relationship between equal group amplification and individual amplification parity}
In this section we discuss the relationship between comparing groups based on group amplification or the distribution of individual amplification values. It is easy to see that if amplification $a(G)$ of a group $G$ were a linear function of the amplification of individuals $i \in G$, that is, when $a(G) = c_1 \cdot \sum_{i \in G} a(i) + c_2$, then individual amplification parity (\eqref{eqn:individual_amplification_parity}) implies equal group amplification (\eqref{eqn:equal_group_amplification}).

However, our definition of amplification does not satisfy this requirement. To see why, consider the function $f(G) = |U_{T_{G}}|$, where $T_G$ is the set of Tweets authored by members of the group $G$ and $U_{T_{G}}$ is the set of users who registered an impression event with at least one Tweet in $T_G$. The function $f$ is a submodular set function exhibiting a diminishing return property $f(G \cup H) \leq f(G) + f(H)$. Equality would hold if Tweets from groups $G$ and $H$ reach completely non-overlapping audiences. For most of the groups we consider, such as groups of politicians from a political party, this is highly unlikely. We define amplification as the ratio between two such submodular set functions. As a consequence, equivalence of amplification at an individual level does not imply equivalence at the group level, and vice versa.

\subsubsection{Identifying outliers via leave-one-out analysis}
To find the most significant outliers in each group in terms of amplification, we performed a leave-one-out or jackknife analysis. We calculated group amplification for each group with each one of the members left out. We then selected positive and negative outliers on the basis of how much leaving the member out of the group has changed the aggregate amplification of the rest of the groups. For a member of the group to be identified as an outlier by this process, it has to have a significant audience (so it has a substantial contribution to the overall group amplification) and it has to behave substantially differently than other members of the group.

\section{An overview of the research process and Twitter’s review process for publications}

This team of authors carried out the research reported here with considerable autonomy and independence throughout the research process. Our findings are shared without cherry picking. In this section we give a detailed account of our research process and steps taken to minimize the influence of corporate interests.

\subsection{Hypothesis selection}

Shortly after the idea of this research was conceived of (in November 2019), the authors have pre-registered a number of hypotheses which could be tested using the massive-scale experiment reported on in this paper. This pre-registration was informal, inasmuch as the pre-registration document was not made public and was not required or reviewed as part of the internal processes at Twitter. These hypotheses centred around algorithmic amplification of abusive content/hate speech, political polarization, and misinformation. No one except the authors of this paper were involved in selecting hypotheses or designing research methods.

Initially the researchers worked in parallel on testing hypotheses related to abusive Tweets and political content. In June 2020, the authors decided to prioritise research on the questions related to political content, because of the higher quality third party data that was available. There were no other considerations that influenced hypothesis selection.

\subsection{Project prioritisation}
The research project was submitted to be prioritised as a research project within the Cortex Applied Research team in December 2019. This prioritization process is used to allocate headcount to research projects. This was the first time the aims and scope of this research project was disclosed widely within the company. The project was judged to be high-impact for the company, and the resulting decision was to allocate headcount, allowing the coauthors to spend more time on the project.

\subsection{External input to the research project}

Consultation with parties outside of the immediate research team (authors of this paper) changed the direction of research in two specific ways:

\begin{itemize}
    \item \textbf{More global scope:} Initial results focused on the United States and United Kingdom. The public policy team pointed out the bias in choosing to focus on these two countries while Twitter is used globally. Following up on this observation we extended the analysis to further countries, and we believe we included all countries where it was technically possible. An effort was made to reduce subjectivity in choosing which countries to include in our analysis: we have extended the analysis to all countries where we could reliably identify a large number of legislators, and where we had enough data from users. The resulting selection criteria are deteiled in SI Section 1.\ref{sec:countries_selection}.
    
    \item \textbf{Using multiple media bias rating datasets:} Initially, the team had access to only one media bias rating dataset (AllSides). Multiple teams raised concerns about the reliability of this third party source, and about making our findings dependent on the validity of a single underlying dataset. We have therefore obtained a license for a second media bias dataset (Ad Fontes Media). We present both sets of results in our paper with no normative judgment on the validity of either of these underlying sources of data.
\end{itemize}

\subsection{Internal review process}
Before a research paper is submitted to peer review (or is published in other form), an internal review is conducted where representatives of the following teams have to give approval:

\begin{itemize}
    \item \textbf{The public policy team} reviews the paper’s likely impact on public policy.
    \item \textbf{The investor relations team} reviews the paper to ensure that it complies with regulations regarding the release of material non-public information.
    \item \textbf{The intellectual property legal team} reviews if the paper discloses any non-public information relating to company IP, and whether any third-party data or intellectual property is used under correct licensing terms.
    \item \textbf{The communications team} reviews the paper to ensure the manuscript is consistent with broader Twitter terminology, and to be able to plan additional communication such as blog posts and Tweets from company accounts. Suggesting changes to the interpretation or presentation of findings in the paper would be beyond the scope of this review.
    \item \textbf{Two technical reviewers}, usually employees of the company and nominated by the authors, comment on the paper’s technical contributions, scientific rigour and the quality of presentation.
\end{itemize}

For this manuscript the internal review started on 27 Aug 2020. None of the reviewers suggested any changes to the paper or made any substantial comments beyond commenting positively on the work or stating their approval. After necessary approvals were given and a full Data Protection Impact Assessment was completed, the manuscript was first shared by a journal editor in a pre-submission enquiry on 25 Sep 2020. While it is possible through this review process for Twitter to refuse a request to publish a paper on various grounds, to our knowledge this has never happened at Twitter.

\subsection{Privacy and Data Protection Reviews}
As the project involved using personal data, several privacy and data protection (PDP) reviews were conducted at various stages to ensure compliance with Twitter's policies regarding data retention and protection of user privacy. These reviews did not alter the research decisions the team made.

\subsection{Ethical Approval}

The control group assessed was not created for the purpose of research but rather for the business purpose of improving the algorithm and providing a baseline by which it could be compared to monitor the ongoing performance of the algorithm. As such, this work was reviewed by Twitter’s legal and privacy teams as part of its ordinary business operations (and not an IRB). As part of this review, a data protection impact assessment was conducted, and it was determined that additional notice and consent mechanisms were not required.

\section{Data availability}
To allow reproduction and limited extension of our findings we make the following data available upon request.

\subsection{Terms of data access} Data will be made available upon request, by emailing the corresponding author. Prior to accessing the data, researchers will be required to sign up for a Twitter Developer Account at  \href{https://developer.twitter.com}{developer.twitter.com}. Access to the Data Set is governed by the \href{https://developer.twitter.com/en/developer-terms/agreement-and-policy}{Twitter Developer Agreement and Policy} available online and prior to accessing the data set all researchers must agree to this agreement. Further sharing of data is not permitted.

\subsection{Details of the dataset}

In this section we give details of data we will make available upon request which allows the reproduction of main findings presented in Figures 1A and 1B, Figure 2, supplementary Figure \ref{fig:party_amplification}. In addition to the comma separated data files whose details are given below, a python jupyter notebook will be provided to reproduce our specific plots from the data.

\texttt{parties\_aggregate\_amplification\_bootstrap.csv} allows the reproduction of Figures 1A, 1B and \ref{fig:party_amplification}, contains amplification values for each political party in one of the 7 countries studied. To increase the robustness of our findings we used bootstrap: each ‘bootstrap fold’ in this dataset relates to different random subset of politicians from each party, with amplification calculated over a random sample of days from within the study period. This file has 7,074 rows and the following columns:

\begin{itemize}[noitemsep,topsep=0pt]
    \item \texttt{grouping\_id}: encodes a legislature and particular way of grouping legislators where ambiguous.
    \item \texttt{group\_label}: identifies a party or political group.
    \item \texttt{bootstrap\_fold\_id}: identifies the bootstrap fold
    \item \texttt{amplification\_ratio}: numerical amplification value expressed as a percentage, as reported in Figures 1 and 2. $0\%$ indicates no relative amplification compared to control condition.
\end{itemize}

\texttt{media\_bias\_categories\_daily\_audience\_bootstrap.csv} allows reproduction of Figure 2, contains the aggregate daily US audience of Ad Fontes or AllSides media bias categories between 15 April and 15 August 2020 within the two experimental conditions. This file has 185,501 rows the following columns:

\begin{itemize}[noitemsep,topsep=0pt]
    \item \texttt{rating\_source}: either adfontes or allsides
    \item \texttt{media\_bias\_category}: identifies media bias category, such as 'Lean Left'
    \item \texttt{date}
    \item \texttt{bootstrap\_fold\_id}
    \item \texttt{experimental\_condition}: either ranking or control
    \item \texttt{audience}: unique number of users within experiment condition who encountered at least one Tweet with a link from this category
\end{itemize}

\texttt{media\_sources\_daily\_audience.csv}: contains more fine grained data on individual U.\,S.\ media sources, not aggregated by media bias ratings but instead listed for each individual publication. This file can be used to reproduce the outliers in Figure 2 of the main paper, and can be used for further analysis and extensions. To mitigate privacy risks associated with the readership of very small publications, we only include the top 150 media outlets we studied, these have large enough audiences to mask any individuals: the smallest daily audience measurement in this CSV is 102. This file has 30,005 rows and the following columns.
\begin{itemize}[noitemsep,topsep=0pt]
    \item \texttt{publication\_title} e.\,g.\ AP or Al Jazeera
    \item \texttt{date}
    \item \texttt{experiment\_condition} either ranking or control, identifies whether the audience is within the treatment or control group
    \item \texttt{audience} number of unique users in each experimental condition who have encountered at least one link from this publication/media source in the given day
\end{itemize}

\texttt{legislator\_screennames.csv} and \texttt{legislator\_screennames\_grouped.csv} (a processed version of the same) contain the list of Twitter accounts of legislators in 7 countries with their associated party affiliations at the time of the analysis. This dataset is derived from publicly available sources and has been improved via manual data curation (see SI 1.\ref{sec:legislators}). While this is not needed to reproduce our results, sharing this data could be useful in validating the quality of the data underlying our analysis and to check details of each legislator's assumed party membership. This file has the following columns:

\begin{itemize}[noitemsep,topsep=0pt]
    \item \texttt{country\_code}
    \item \texttt{chamber} e.g. US House vs Senate
    \item \texttt{group\_id} name of political party or group
    \item \texttt{screenname}
\end{itemize}

\texttt{media\_sources\_regex.csv} contains regular expressions we used to extract Tweets containing a link to one of the US media sources we analyzed (See SI 1.\ref{sec:regex}). This is not needed to reproduce any figures, but allows additional scrutiny of our methods. This file has the following columns:

\begin{itemize}[noitemsep,topsep=0pt]
    \item \texttt{media\_source\_name}
    \item \texttt{media\_source\_domain} the domain name associated with the online news source
    \item \texttt{is\_opinion} binary value indicating whether this row relates to editorial/opinion content or regular news reporting. Used to distinguish between e.\,g.\ NY Times News and NY Times opinion as separate media sources.
    \item \texttt{regex} regular expression matched against a url to determine if the URL points to an article from the media source in question
\end{itemize}

\subsection{Third party media bias data}

Licensing terms do not permit us to share raw data from Ad Fontes Media or AllSides which we used in our analysis. Researchers who wish to reproduce the mapping of media sources to media bias categories are encouraged to obtain a license and the data from the AllSides and Ad Fontes Media websites.

\begin{figure}
\centering
\includegraphics[width=\textwidth]{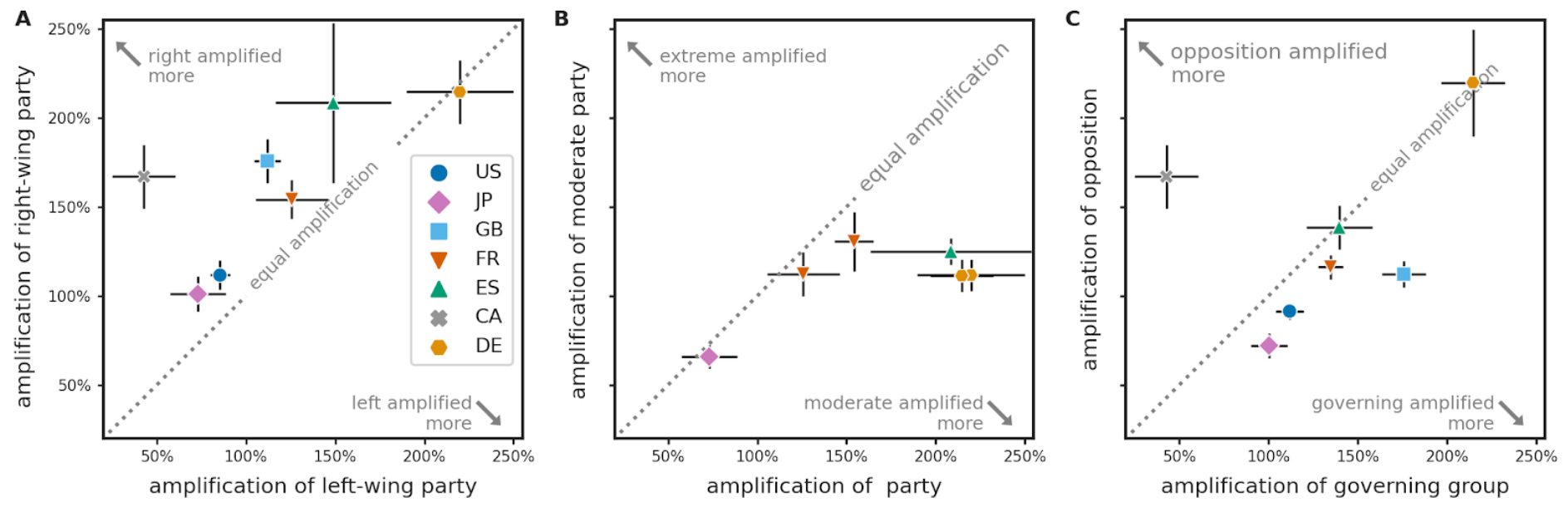}
\caption{Evaluating various hypotheses about algorithmic amplification of political parties. \textbf{A} (This panel is identical to Fig. 1B) Comparing the amplification of the largest mainstream left- and right-wing parties in each country: Democrats vs Republicans in the U.S., CDP vs LDP in Japan, Labour vs Conservatives in the U.K., Socialists vs Republicans in France, PSOE vs Popular in Spain, Liberals vs Conservatives in Canada and SPD vs CDU/CSU in Germany. \textbf{B} Comparing extreme far-left or far-right parties against relevant mainstream parties from the same country: CDP vs JCP (left) in Japan; LFI vs Socialists (left) and RN vs Republicans (right) in France; VOX vs Popular (right) in Spain; Die Linke vs SPD (left) and AfD vs CDU/CSU (right) in Germany. \textbf{C} Comparing the governing parties against the opposition parties in each country. In the United States Republicans were considered governing. Error bars in all panels show standard error estimated from bootstrap.}
\label{fig:party_amplification}
\end{figure}

\begin{figure}
\centering
\includegraphics[width=\textwidth]{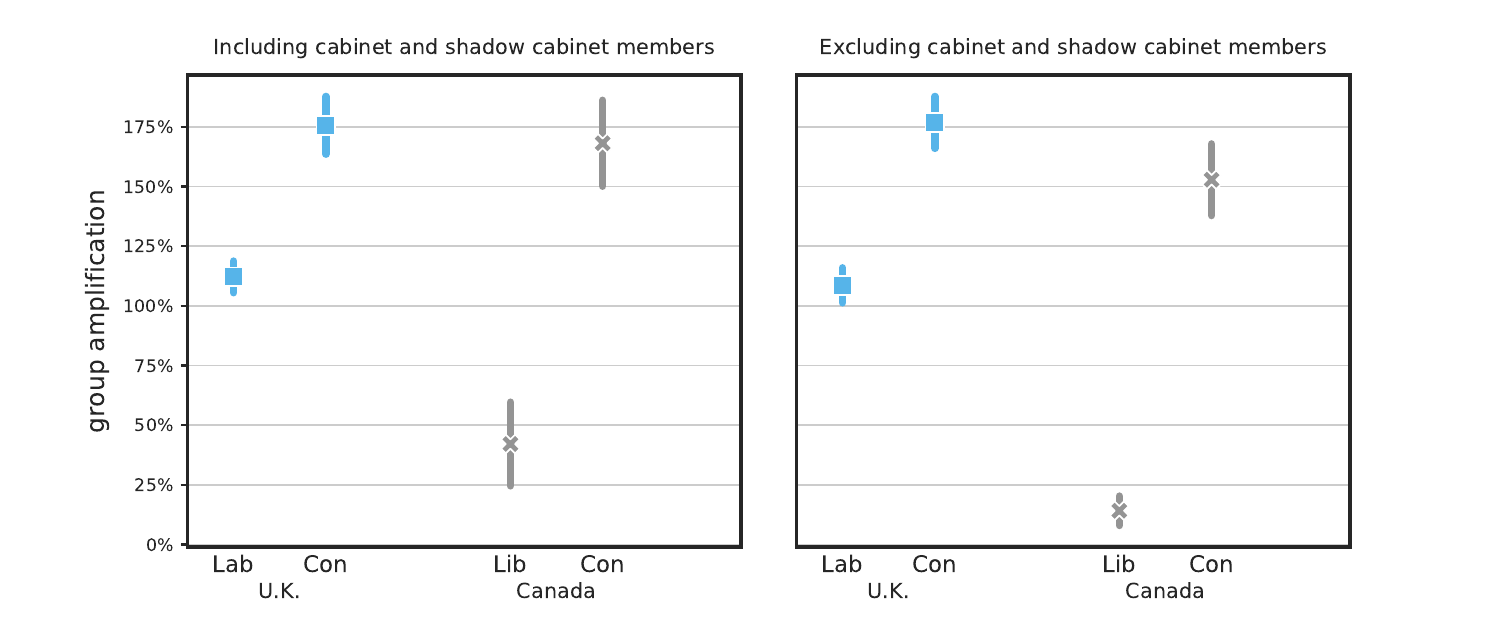}
\caption{Ablation study reproducing findings from Figure 1 for the U.\,K.\ and Canada with members of the cabinet and shadow cabinet removed from the analysis. \emph{Left panel:} In these countries the Prime Minister and key members of the government are also elected Members of Parliament, they are thus included when calculating group amplification of each party in Figure 1A. \emph{Right panel:} We rerun the analysis with members of the Cabinet and members of the Shadow Cabinet (high ranking opposition politicians) left out. Inthe U.K. we excluded Prime Minister Boris Johnson, and holders of Great Offices of the State Matt Hancock, Rishi Sunak,  Priti Patel and Dominic Raab. Likewise, we excluded Shadow Prime Minister Keir Starmer and Shadow Cabinet Members Angela Rayner, Lisa Nandy, Anneliese Dodds and Jonathan Ashworth. In Canada we excluded Prime Minister Justin Trudeau and cabinet members Chrystia Freeland, Harjit Sajja, Catherine McKenna and Ahmed Hussen and opposition politicians Andrew Scheer, Pierre Poilievre, Erin O'Toole and Candice Bergen. Our findings are qualitatively similar irrespective of whether cabinet members are included or not.}
\label{fig:party_cabinet_ablation}
\end{figure}

\begin{figure}
\centering
\includegraphics[width=0.7\textwidth]{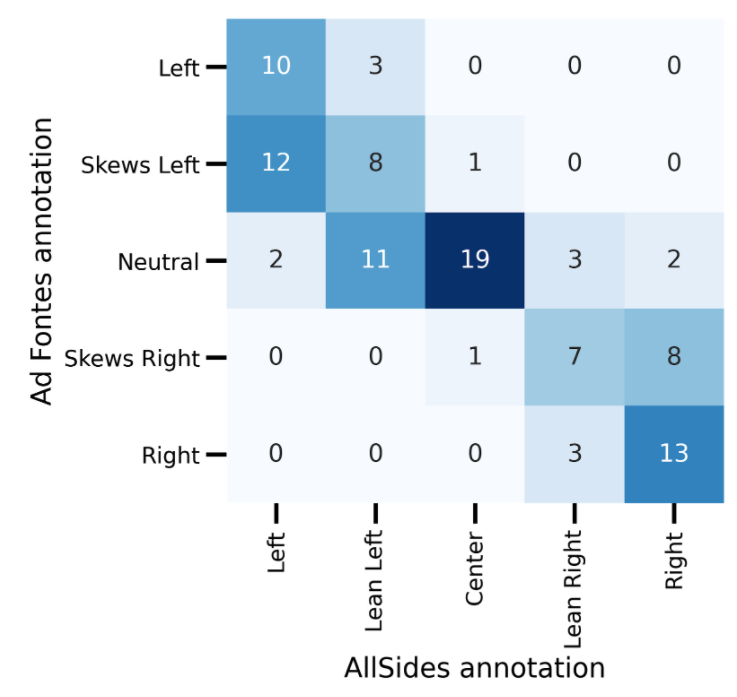}
\caption{Number of news sources for a particular combination of AllSides and Ad Fontes rating, for publications where both ratings were available. The comparison reveals a tendency for Ad Fontes to rate publications more neutrally compared to AllSides, especially on the Left end of the spectrum. For example, the majority of sources rated as ‘Left’ by AllSides is rated as ‘Skews Left’ by Ad Fontes. Similarly, the majority of sources rated ‘Lean Left’ by AllSides is rated ‘Neutral’ by Ad Fontes.}
\label{fig:news_sources_count}
\end{figure}

\begin{figure}
\centering
\includegraphics[width=0.7\textwidth]{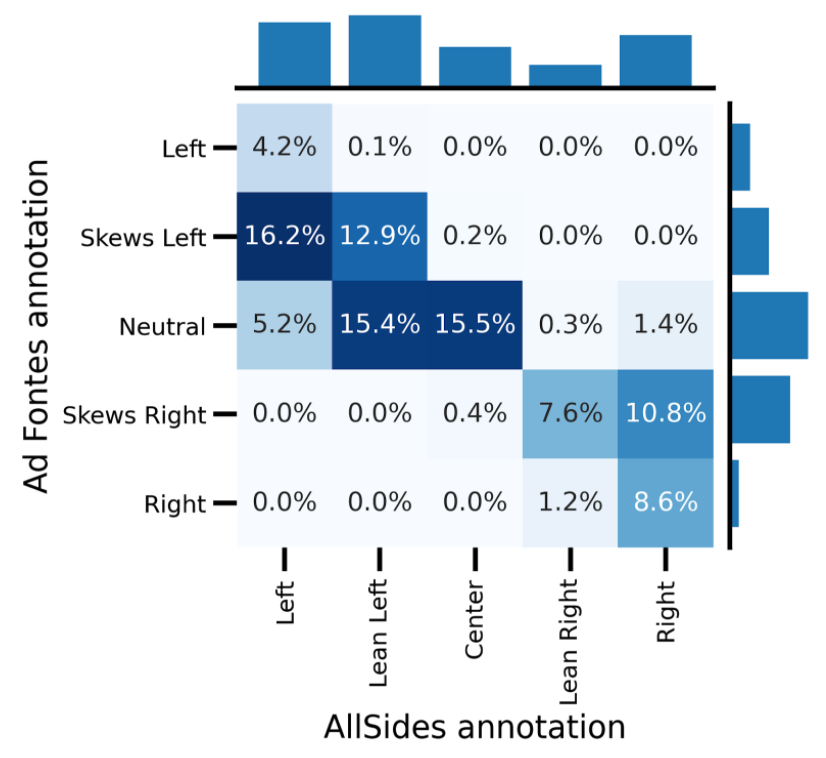}
\caption{Distribution of AllSides and Ad Fontes media bias ratings among Tweet impressions in the control group. The annotated heatmap shows the percentage of all Tweet impressions with a particular combination of AllSides and Ad Fontes rating. The most frequently occurring combination is AllSides Left, Ad Fontes Skews Left, accounting for 16.2\% of Tweet impressions. Histograms show the marginal distribution of AllSides and Ad Fontes annotations. The most frequent Ad Fontes category is Neutral, while the most frequent AllSides category is ‘Lean Left’.}
\label{fig:news_sources_impressions}
\end{figure}

\begin{table}\centering
\caption{Most popular news sources for each combination of AllSides and Ad Fontes media bias rating by number of Tweet impressions in the control group.}

\begin{tabular}{llr}
Ad Fontes & AllSides & Top media sources by number of impressions \\
\midrule
Left & Left & Daily Beast, Slate, Intercept \\
Left & Lean Left & Truthout, FAIR.org, The Advocate \\
Skews Left & Left & CNN (opinion), BuzzFeed, Vox, Raw Story\\
Skews Left & Lean Left & CNN (news), Politico, NBC \\
Skews Left & Center & The Week \\
Neutral & Left & New York Times (opinion), NY Daily News \\
Neutral & Lean Left & New York Times (news), Washington Post, LA Times \\
Neutral & Center & Bloomberg, Business Insider, Associated Press \\
Neutral & Lean Right & Reason, Marketwatch, Fiscal Times \\
Neutral & Right & New York Post, Daily Mail \\
Skews Right & Center & Real Clear Politics \\
Skews Right & Lean Right & Fox News (news), Washington Examiner, Washington Times \\
Skews Right & Right & Fox News (opinion), National Review, The Epoch Times \\
Right & Lean Right & One America News Network, PJ Media, Judicial Watch \\
Right & Right & Breitbart, Daily Caller, The Gateway Pundit \\
\bottomrule
\end{tabular}
\label{tab:ad_fontes_allsides_media}
\end{table}

\FloatBarrier

\bibliography{supplementary}